\documentclass[traditabstract]{aa} 
\usepackage{graphicx}
\usepackage{txfonts}

\newcommand{\CP}[3]   {\mbox{Computers in Physics~\textbf{#1},~#2~(#3)}}

\newcommand{\ApJ}[3]    {\mbox{ApJ~\textbf{#1},~#2~(#3)}}
\newcommand{\ApJS}[3]   {\mbox{ApJ~Suppl.~\textbf{#1},~#2~(#3)}}
\newcommand{\ApJL}[3]   {\mbox{ApJ~Lett.~\textbf{#1},~#2~(#3)}}

\newcommand{\AJ}[3]     {\mbox{Astron.~J.~\textbf{#1},~#2~(#3)}}
\newcommand{\MNRAS}[3]  {\mbox{MNRAS~\textbf{#1},~#2~(#3)}}

\newcommand{\astroph}[1]{\mbox{\texttt{astro-ph/#1}}}

\begin{document}

\title{ASOHF: a new adaptive spherical overdensity halo finder\\}

\author{Susana Planelles\thanks{susana.planelles@uv.es}  
\and Vicent Quilis\thanks{vicent.quilis@uv.es}}

\institute{Departament   d'Astronomia    i   Astrof\'{\i}sica,   
Universitat   de
Val\`encia,   46100   -    Burjassot   (Valencia),   Spain\\}

\date{Received date / Accepted date}

\abstract  {We present and test a  new  halo  finder based  on  the  spherical
overdensity (SO)  method. This  new adaptive
spherical  overdensity halo finder  (ASOHF) is  able to  identify dark
matter haloes  and their substructures (subhaloes) down  to the scales
allowed  by the  analysed simulations.   The code  has  been especially
designed for the adaptive mesh refinement cosmological codes, although
it can be used  as a stand-alone halo finder for  N-body codes. It has
been  optimised for the  purpose of  building the  merger tree  of the
haloes. In  order to verify  the viability of  this new tool,  we have
developed a set of bed tests that allows us to estimate the performance
of the  finder.  Finally, we apply  the halo finder  to a cosmological
simulation and  compare the results  obtained to those given  by other
well known publicly available halo finders.}


\authorrunning{S. Planelles and V. Quilis}

\titlerunning{ASOHF: a new adaptive spherical overdensity halo finder}

\maketitle

\section{Introduction}

In the  standard model of structure formation,  small systems collapse
first   and   then  merge  hierarchically   to   form   larger
structures. Galaxy clusters,  which are at the top  of this hierarchy,
represent the  most massive virialized structures in  the universe and
may host thousands of galaxies.

Numerical simulations  of structure  formation are essential  tools in
theoretical cosmology.  During the  last years, these simulations have
become a  powerful theoretical mechanism to  accompany, interpret, and
sometimes to lead cosmological  observations because they bridge the gap
that often  exists between basic  theory and observation.   Their main
role, in addition  to many other uses, has been  to test the viability
of the different structure formation models, such us, variants
of the  cold dark matter  (CDM) model, by evolving  initial conditions
using basic physical laws.

Historically, the use of cosmological simulations started in the 1960s
(\cite{Aarseth}) and 1970s  (e.g., \cite{Peebles} and \cite{white76}).
These  calculations were N-body  collisionless simulations  with few
particles.  Over the  last three decades great progress  has been made
in  the development  of N-body  codes that  model the  distribution of
dissipationless  dark   matter  particles.   Besides   this  numerical
progress, computers and computational resources have made such
progress  that  simulations  could be  applied  systematically  as
scientific  tools.  Their  use  led  to  important results  in  our
knowledge of the Universe.

In addition  to the treatment  of the collisionless dark  component of
cosmological structures, hydrodynamical codes designed to describe the
baryonic component  of the Universe have also been  developed, usually
coupled with N-body codes.

The generation of the data is  only a first step that carries
out complex simulations to generate  a huge amount of raw information.
In  the  particular case  of  N-body  simulations,  the aggregates  of
millions  of dissipationless  dark  matter particles  produced in  the
simulations require  to be interpreted  and somehow compared  with the
observable Universe.  To do so, it  is necessary to identify
the groups  of gravitationally bound  dark matter particles,  which are
the dark counterparts of the observable components of the cosmological
structures (galaxies, galaxy clusters, etc.).  These dark matter clumps
are the so-called dark matter haloes,  and the task to identify them in
simulations is  usually carried out  with the help of  numerical tools
known as halo finders.

There are several kinds of  halo finders currently widely used.  All
these halo-finding algorithms seem  to perform exceedingly  well when
they  deal with  the  identification of  haloes without  substructure.
However, the remarkable  development of N-body simulations
and the  applications studied  with these new  codes necessitated   
new  algorithms  able   to  deal  with  the   scenario  of
haloes-within-haloes   (\cite{klypin99a};       \cite{klypin99b};
\cite{moore99}).

Motivated  by the  importance of  working  with the  halo and  subhalo
population obtained from  simulations, we present  a new
halo finder called ASOHF (adaptive spherical overdensity halo finder).
This finder, based on the  spherical overdensity (SO) method, has been
especially designed  to couple  with the outputs  of the  adaptive mesh
refinement (AMR) cosmological Eulerian codes.  Taking advantage of the
capabilities of the AMR scheme,  the algorithm is able to identify  
dark  matter  haloes  and  their  substructures
(subhaloes) in a  hierarchical  way limited only by the 
resolution of the analysed simulation.
Additionally,  it has  been  optimised for  building the  evolutionary
merger tree  of the haloes.  To check the  viability of this
new  tool, we developed a  simple set  of tests  which generates
different toy models with properties completely known beforehand. This
battery of tests  will allow us to calibrate the  real accuracy of our
finder.  Finally, we  apply ASOHF to a cosmological  simulation and
compare the results with  those obtained with other publicly available
halo-finding  algorithms, such  as AFoF  (\cite{vanKampen95})  and AHF
(\cite{Knollmann}).

The present  paper is organized as  follows.  In Sect.  2 we briefly
describe several existing halo-finding  algorithms tested
in  the scientific  literature.  In  Sect. 3  we introduce  the main
properties of the halo finder ASOHF.  
In Sect. 4 we describe several
idealised tests and the results of applying ASOHF to them.  In Sect.
5  ASOHF is  applied to  a  cosmological simulation,  and the  results are
compared  to those  obtained by  other halo  finders available  in the
literature.   Finally, in  Sect.  6, we  summarize  and discuss  our
results.

\section{Background}

Different  algorithms  to  identify  structures and  substructures  in
cosmological  simulations  have  been  proposed  and  have  seen  many
improvements  over the  years.  As  a consequence,  there  are several
kinds of  halo finders currently  wide used although, at  heart, the
basic idea of them all is to identify gravitationally bound objects
in  an N-body  simulation.  Let  us briefly outline  some  of  the most
popular halo finders.

The  classical method  to identify  dark matter  haloes is  the purely
geometrical friends-of-friends algorithm (FoF) (\cite{Davis85}).  This
technique consists  in finding neighbours of dark  matter particles and
neighbours  of these  neighbours  according to a given linking  length
parameter.  The  characteristic linking length,  $l_{FoF}$, is usually
set to $\sim  0.2$ of the mean particle  separation. The collection of
linked  particles forms  a group  that is  considered as  a virialized
halo. The mass  of the halo is defined  as the sum of the  mass of all
dark matter particles within  the halo.  Among the main advantages
of this  algorithm we  can point out  that its results  are relatively
easy  to  interpret  and  that  it does  not  make  any  assumption
concerning  the   halo  shape.   The  greatest   disadvantage  is  its
rudimentary choice of linking length, which can lead to a connection of
two separate objects via the so-called ''linking bridges''.  Moreover,
because  structure   formation  is  hierarchical,   each  halo  contains
substructure  and  thus,  different  linking  lengths  are  needed  to
identify   ``haloes-within-haloes''.   There   are   several  modified
implementations of the  original FoF, such as the  adaptive FoF (AFoF;
\cite{vanKampen95}) or the  hierarchical FoF (HFoF; \cite{klypin99a}),
among others, which try to overcome these limitations.

The   other  most   general  method   is  the   spherical  overdensity
(SO, \cite{lacey94})  that uses  the  mean overdensity  criterion for  the
detection of virialized haloes. The basic idea of this technique is to
identify spherical  regions with  an average density  corresponding to
the  density of a  virialized region  according the  top-hat collapse.
The main  drawback of  the SO  mass definition is  that it  is somehow
artificial,  enforcing spherical  symmetry  on all  objects, while  in
reality haloes often have an irregular structure (\cite{white02}), for
example, haloes that were formed in a recent merger event or haloes at
high redshifts.   Furthermore, defining an  SO mass can  be ambiguous
because the corresponding SO spheres might
overlap for two close density  peaks.   
Due  to  these   characteristics,  the  SO  method  implies
oversimplifications  that  could lead  to  unrealistic results and  which
therefore deserve  a  careful  treatment.   Despite these  apparently  significant
disadvantages, one of the most  relevant features of this technique is
that no linking length is needed to define the structures.

Almost  all  existing halo  finders are  based on  either  the FoF
algorithm, the SO, or a combination of both methods.
  
The  DENMAX (\cite{bertschinger91};  \cite{gleb94})  and   the  SKID
(\cite{Weinberg97}) algorithms  are similar  methods.   Both of  them
calculate  a  density  field  from  the  particle  distribution,  then
gradually move  the particles  in the direction  of the  local density
gradient  ending with  small  groups of  particles  around each  local
density maximum.  The FOF method is then used to associate these small
groups with individual haloes.  The difference between the two methods
is in the calculation of the  density field.  DENMAX uses a grid, while
SKID applies an adaptive smoothing  kernel similar to that employed in
SPH  techniques (\cite{gingold};  \cite{lucy}).  The  effectiveness of
these  methods is  limited  by  the technique  used  to determine  the
density field (\cite{gotz98}).
 
The HOP method (\cite{EiHu98} ) is based on a density field similar to
the SKID. However, it uses  a different type of particle sliding.  The
HOP  algorithm searches  for the  maximum  density among  the {\it  n}
nearest  neighbours of  a particle  and  attaches the  particle to  the
densest  neighbour. Finally,  it groups  particles in  a  local density
maximum, defining a virialized halo.

The BDM  method (\cite{klypin99a})  uses randomly placed  spheres with
predefined radius, which are iteratively moved to the centre of mass of
the particles contained in them until the density centre is found.

Completely different is  the VOBOZ technique (\cite{neyrinck05}), which
uses a Voronoi tessellation to calculate the local density.

The  halo  finder  MHF  (\cite{gkg04})  took  advantage  of  the  grid
hierarchy generated by the AMR code MLAPM (\cite{knebe01}) to find the
haloes in a given simulation.   In most of the cosmological AMR codes,
the grid hierarchy is built in such  a way that the grid is refined in
high-density   regions  and  hence   naturally  traces   the  densest
regions.  This can  be used  not only  to select  haloes, but  also to
identify the substructure.  The AHF  (Amiga Halo Finder), which is the
direct successor of the original  MHF, has been recently presented and
tested (\cite{Knollmann}).

Similarly to SKID, in  SUBFIND (\cite{springel01}) the density of each
particle is estimated with a  cubic spline interpolation.  In a first
step, the  FOF method is used  and then any  locally overdense region
enclosed by  an isodensity  contour that traverses  a saddle  point is
considered  as a  substructure candidate.  SUBFIND runs  on individual
simulation snapshots,  but can afterwards reconstruct  the full merger
tree of each  subclump by using the subhalo  information from previous
snapshots.  All  subhalo  candidates  are then  examined  and  unbound
particles are removed.

A   quite   different   method   is   provided  by   the   code   SURV
(\cite{tormen04},     \cite{giocoli08},    \cite{giocoli09}),    which
identifies subhaloes  within the  virial radius of  the final  host by
following all  branches of the merger  tree of each  halo (rather than
just the main  branch), in order to reconstruct  the full hierarchy of
substructure down to the mass resolution of the simulation.

The parallel  group finder 6DFOF  (\cite{diemand06}; \cite{diemand07})
finds  peaks in  phase-space density,  i.e., it  links the  most bound
particles inside the cores of  haloes and subhaloes together. The same
objects identified  by 6DFOF at different times  therefore always have
quite a large fraction of particles  in common (in most cases over $90
\%$). This makes finding progenitors or descendants rather easy.

\section{The ASOHF halo finder}

The  halo identification is  a crucial  issue in  the analysis  of any
cosmological  simulation.   Inspired   by  this,  we
developed  a halo  finder  especially  suited for  the  results of  the
Eulerian cosmological  code MASCLET (\cite{quilis04}),  although it can
be easily applied to the outcome of a general N-body code.

The  halo finder developed  for MASCLET,  ASOHF, shares  some features
with AHF (\cite{Knollmann}), which is the evolution of the original MHF
halo finder  (\cite{gkg04}).  Although we used an identification
technique based on  the original idea of the  SO method, the practical
implementation of our finder has several steps designed to improve the
performance of this method and to get rid of the possible drawbacks as
well as to take advantage of an AMR grid structure.

The particular implementation of  our halo finder follows several main
steps.
  
\begin{enumerate}

\item
In a first  step, the algorithm reads the density  field computed on a
hierarchy of grids provided by the simulations. Then the SO method is
applied to  each density  maximum: radial shells  are stepped  out around
each  density peak  until the  mean  overdensity falls  below a  given
threshold or  there is  a significant  rising in  the slope  of the
density profile.  The overdensity,  $\Delta_c$, depends on the adopted
cosmological model and can be approximated by the expression
(\cite{brynor98})
\begin{equation}
\Delta_c= 18 \pi^2 + 82 x - 39 x^2 ,
\end{equation}
\noindent
where  $x=\Omega(z)-1$  and  $\Omega(z)=[\Omega_m (1+z)^3]/  [\Omega_m
(1+z)^3  +  \Omega_\Lambda]$.  Typical  values  of  this contrast  are
between 100 and 500, depending on the adopted cosmology.

Therefore, the  virial mass  of a halo,  $M_{vir}$, is defined  as the
mass enclosed  in a spherical  region of radius, $r_{vir}$,  with an
average density $\Delta_c$ times  the critical density of the Universe
$\rho_c(z)=3H(z)^2/8\pi G$
\begin{equation}
M_{vir} (\leq r_{vir} )=\frac{4}{3}\pi \Delta_c\rho_c r_{vir}^3 .
\end{equation}

This first  step, which only defines  the scale of the  objects we are
looking for, provides  a crude estimation of the  position, radius and
mass for each detected halo.

\item
The second step takes care  of possible overlaps among the preliminary
haloes found in  the first step. 
In our method,  if two haloes overlap
and the shared mass is larger than the 80\% of the minimum mass of the
implicated haloes, the less massive  of them is removed from the list.
On the other hand, if the shared mass is between the  40\% and the 80\% 
of the minimum mass of the haloes,
the algorithm  joins these haloes and computes the  centre of
mass of the new halo.  Consequently, it removes the less massive halo
from the list,  and applies again the first step to  the new centre of
mass to obtain the  physical properties of the new halo. 
In the end, this step provides a final number of haloes.
  
 \item
Once we have a tentative halo  selection, a third step provides a more
accurate sample by working only  with the dark matter particles within
each halo.  These particles  are distributed through the complete simulated
volume and  are not limited by  cell boundaries.  ASOHF  can deal with
several particles species (particles with different masses), providing
therefore a best-mass resolution.  This step is  crucial  to
obtain a  precise estimation  of the main  physical properties  of the
haloes, particularly, a  new prediction for the mass  and position of the
centre of mass.

\item
Once centres  of potential  haloes are found,  the code checks  if all
particles  contained  in  a  halo  are  bound.   It  finds  the  final
properties and the radial structure of  all haloes in the same way
as in  the first step,  but now working  only with the  particles. It
places  concentric shells  around each  centre and  for each  shell it
computes the mass of the dark matter particles, the mean velocity, and
the velocity  dispersion relative to  the mean. In order  to determine
whether  a particle is  bound or  not, the  code estimates  the escape
velocity at the position  of the particles (\cite{klypin99a}).  If the
velocity  of  a particle  is  higher  than  the escape  velocity,  the
particle is assumed to be  unbound and is therefore removed from the
halo considered. This pruning is  halted when a given halo holds fewer
than a fixed minimum number  of particles or when no more particles
need to be removed.

\item
The process finishes when it  verifies that the radial density profile
of the haloes  is consistent with the functional  form proposed by NFW
(\cite{nfw97})  in  the  range  from  twice the  force  resolution  to
$r_{vir}$

\begin{equation}
\rho (r)= \frac{\rho_0}{(r/r_s)^{\alpha} [1+(r/r_s)]^{\beta}} ,
\end{equation}

where $\rho_0$ is the normalization, $\alpha$ and $\beta$ are the
inner and outer slopes respectively, and $r_s$ is the scale radius.
The virial  and the scale radius are related through  
the concentration parameter ${\it C}={r_{vir}/r_s}$.

\end{enumerate}

After  this process,  the final  output for  each halo  includes a
precise estimation of  its main physical properties,  the list of its
bound particles, the location and  velocity of its centre of mass, and
the density and velocity radial profiles.

Note  that this method is  completely general and
easily applicable  to any N-body  code, assuming the density  field is
previously evaluated on a grid or set of nested grids.

\subsection{Substructure}

Substructures within haloes are  usually defined as locally overdense
self-bound particle groups identified within a larger parent halo.

In our  analysis, the  process of halo-finding outlined above  can be
performed  independently   at  each  level  of   refinement  of  the
simulation.  Then   our  halo  finder  can  trace
haloes-in-haloes in  a  natural  way obtaining a hierarchy  
of nested haloes. Moreover, it
is able  to find several  levels of substructure  within substructure.
This  property  allows  us  to  take advantage  of  the  high  spatial
resolution provided  by the AMR  scheme, identifying a wide  variety of
objects with very different masses and scales.

Still, due to this procedure and to the nature of the AMR grid, this
technique  could  mix   real  substructures  and  overlapping  haloes.  
In  order to  deal with possible  misidentifications of
subhaloes, we need to implement an extra mechanism.

Let  us  consider  two  haloes  from  two  different  but  consecutive
refinement  levels,  $h_1$  (lower  level  and, hence,  lower
resolution)  and $h_2$  (upper  level and,  therefore, higher  spatial
resolution), with masses  $m_1$ and $m_2$, radii $r_1$  and $r_2$, and
the velocity of  the centre of  mass equal to  $ {\it v}_{cm1}$  and ${\it
v}_{cm2}$, respectively.   In our method, these  haloes are considered
as host halo (lower level) and subhalo (upper level), respectively, if
they satisfy the conditions

\begin{enumerate}
\item{} $\kappa=min(m_1,m_2)/max(m_1,m_2) \leq 0.2$
\item{}  $distance(h_1,h_2)<1.2 \, r_1$
\item{} $h_2$ gravitationally bound to $h_1$ .
\end{enumerate}
\noindent

Whilst the  first condition ensures that  the halo in  the lower level
(the biggest one) has a reasonable  mass to host the halo in the upper
level,  the second  condition checks  if  the smaller  halo is  placed
within the radius of the biggest  one or at least in its surroundings.
The last condition is essential to guarantee that the subhalo
is gravitationally bound to its host and therefore define the system
halo-subhalo.

On the  other hand, given the  structure of nested  grids generated by
the AMR scheme,  it is possible that sometimes the  same halo would be
identified more  than once in  different levels with  centres slightly
shifted in position. To  capture these duplicates, a similar criterion
to that used for the substructures is used, but now with the conditions

\begin{enumerate}
\item{}  $\kappa=min(m_1,m_2)/max(m_1,m_2) \textgreater 0.3$
\item{}  $distance(h_1,h_2)< min(r_1,r_2)$
\item{}  $\mid {\it v}_{cm1}-{\it v}_{cm2} \mid /min(\mid {\it v}_{cm1}\mid 
,\mid {\it v}_{cm2} \mid) \leq 1$ .
\end{enumerate}

The set of these three conditions checks if two considered haloes have
similar masses,  positions and velocities  of their centres  of mass.  
If two  haloes satisfy  these conditions,  they  are a
duplicate identification of the same  halo, and then the halo from the
upper level is considered as a misidentification of the other halo and
is dropped out of the list.

At this point, substructures are  defined only on the different levels
of the  grid.  These levels have  been defined and  established by the
assumed refining  criteria which can  be fixed by the  evolution, when
the outputs are directly imported from  a code like MASCLET, or by any
other  criteria, like  the number  of particles  per cell,  when ASOHF
works as a stand-alone code. Thus, ASOHF is able to find substructures
and assign masses to them with  a good accuracy throughout most of the
host halo  and is only limited by  the existence of  refinements in the
computational grid.

Once the code  has acted on the different levels  of resolution of the
considered grid, it  obtains a single halo sample  classifying all the
haloes in  three categories according  to their nature:  single haloes
(with or  without significant substructures),  subhaloes (belonging to
single haloes)  and poor haloes (in  our method these  are haloes with
less than a fixed number of dark matter particles, e.g.  50, or haloes
that are  a misidentification of other haloes).  Thus,
it  is possible  to  obtain a  complete  sample of  objects with  very
different masses and  scales, ranging from the biggest  haloes down to
the  minimum  scales  imposed   by  the  resolution  of  the  analysed
simulations.

One of  the main  advantages of  our method is  that the  hierarchy of
nested  grids  used  by  the  AMR cosmological  simulations  is  built
following the  density peaks, and  therefore these grids  are already
suitably adjusted to track the dark matter haloes.  Last but not least
the  use of AMR  grids implies that  we need no longer define a linking length.

\begin{figure}
\begin{center}
\scalebox{0.8}{\includegraphics[width=12cm]{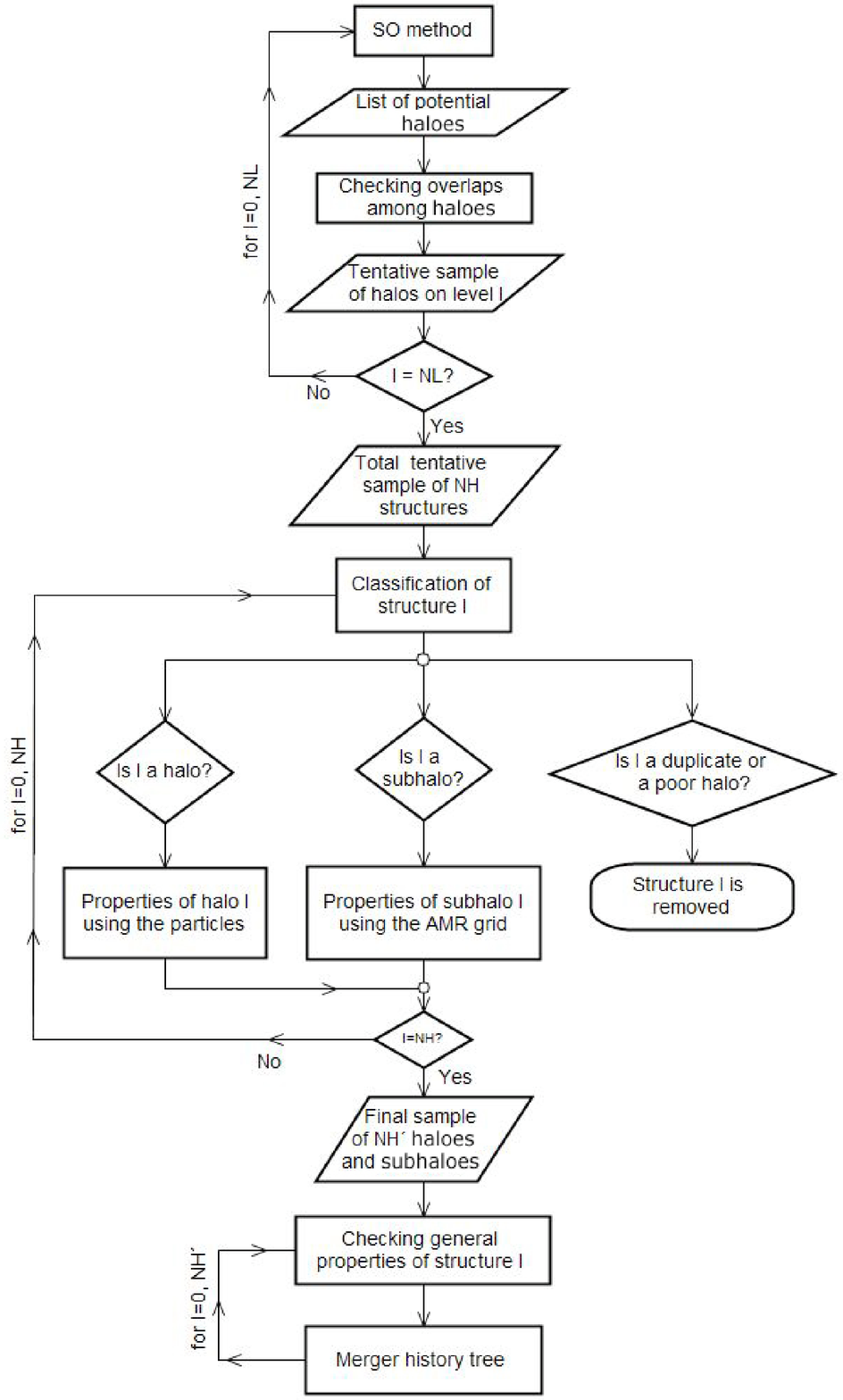}}\\
\end{center}
\caption[Flowchart]{Flowchart for ASOHF. In this diagram, NL stands for
the total number of analysed AMR grid levels. NH ad NH' represent the
total number of tentative and final found haloes, respectively.}
\label{asohf_flowchart}
\end{figure}    

\subsection{Merger tree}

Dark  matter haloes and  their mass  assembly histories  are essential
pieces  of any  non-linear  structure formation  theory  based on  the
$\Lambda$CDM model.   Yet the construction of a  merger tree from
the outputs of an N-body simulation  is not a trivial matter.  We
included in the  halo finder programme a routine that  is able to obtain
the evolution history of each one  of the found haloes.  The method of
progenitor  identification is  based  on the  comparison  of lists  of
particles belonging to  the haloes at different moments  both backwards and
forwards in time,  i. e., it tracks the history of  all dark matter
particles belonging to  a given halo at a  given epoch. This procedure
is  repeated backwards  in  time  until the  first  progenitor of  the
considered halo  is reached.   This mechanism allows  us not  only to
know all progenitors of each halo, but also the amount of mass received
from each one of its ancestors.

This mechanism  can be  applied to build  the merging history  tree of
either the haloes or the subhaloes of the simulation.

This procedure is very useful  when we are interested in an exhaustive
analysis of all  the linking relations among the  haloes, for example,
when  we want to  analyse mergers  or collisions  between two  or more
haloes,  or  when  we  are  interested  in  following  the  history  of
individual haloes as well as different processes of halo disruption.

However, sometimes  we are only interested  in the main  branch of the
merger tree of each halo, or  in other words, in a "simplified" merger
tree.  In  order to have a quick  estimate of the history  of the main
branch  we included a  reduced  merger tree
routine in  the halo  finder 
which, instead  of following all particles  of the haloes,
looks only for the closest particle  to the centre of each halo.  This
particle, which is supposed to be the most bound particle in the halo,
is  followed backwards  in  time  until the  first  progenitor of  the
considered  halo is  found.  With  this  method, each  parent halo  is
allowed to have only one descendant.

The ASOHF method is  summarized in Figure \ref{asohf_flowchart}, where
a flowchart of the main process is shown.

\subsection{Halo shapes}

In ASOHF  code the shape of  the haloes is  evaluated by approximating
their mass distribution  by a triaxial ellipsoid. The  axes of inertia
of the different haloes and subhaloes are evaluated from the tensor of
inertia (see e.g. \cite{cole96}; \cite{shaw06}):
 
\begin{equation}
I_{\alpha \beta}= \frac{1}{N_p} \sum_{i=1}^{N_p} {r_{i, \alpha} r_{i, \beta}}   ,
\end{equation}
where the positions $r_i$ are given with respect to the centre of
mass and the summation is over  all particles in the halo ($N_p$). The axes of
the ellipsoid  can be determined  from the eigenvalues  $\lambda_i$ of
the  inertia tensor as  $a= \sqrt{\lambda_1}$,  $b= \sqrt{\lambda_2}$,
$c= \sqrt{\lambda_3}$, where $a \ge b  \ge c$. Haloes can be classified in
terms  of their  axis ratios  by defining  their degree  of sphericity
\textit{s}, prolateness \textit{q}, and oblateness \textit{p} as

 \begin{equation}
 s=\frac{c}{a};   p=\frac{c}{b};   q=\frac{b}{a} .
 \end{equation}

An additional  measure for the shape of the ellipsoid is the 
triaxiality parameter (\cite{franx91}),

\begin{equation}
T=\frac{a^2-b^2}{a^2-c^2} .   
\end{equation} 
  
An ellipsoid is considered oblate if $0<T<1/3$, triaxial with 
$1/3<T<2/3$, and prolate if $2/3<T<1$.

\section{Testing the halo finder}

Before using  the ASOHF finder  in real cosmological  applications, we
have to  be sure that it  provides accurate and  credible results.  In
order to  validate and  assess the robustness  of our method,  we 
developed  a  set  of  tests  that  will  allow  us  to  quantify  the
uncertainty of the  halo finder algorithm and to  check the properties
of the haloes found with it.

In  these  tests  we  build mock  distributions  of  dark  matter
particles,  made  by  hand,  resembling real  outputs  of  cosmological
simulations.  Therefore, we have perfectly known distributions of dark
matter  particles forming  a given  number of  cosmological structures
with  physical parameters completely  known. Once  these distributions
are built,  we apply the halo  finder and  compare the results
obtained with the  ones originally adopted to create   the mock
distributions by hand.

The different numerical implementations presented in this section 
were performed assuming  the following cosmological parameters: matter
density    parameter,    $\Omega_m=0.25$;    cosmological    constant,
$\Omega_{\Lambda}=\Lambda/{3H_o^2}=0.75$;  baryon  density  parameter,
$\Omega_b=0.045$;    reduced   Hubble   constant,    $h=H_o/100   km\,
s^{-1}\,Mpc^{-1}=0.73$;  power  spectrum  index,  $n_s=1$;  and  power
spectrum normalization, $\sigma_8=0.8$.

The  set of  tests was designed  to help  us  check different
aspects of interest in cosmological  simulations: i) test 1 and test 2
are  focussed  in looking  for  and  characterizing  single haloes  and
subhaloes, respectively, ii) test 3 builds the merger trees of haloes,
and iii) test 4 analyses big samples of haloes.

All  cases considered  in  this section  were  placed in  a
comoving volume of $100 \, h^{-1}\, Mpc$ on a side.  The computational
box has been discretised with  $256^3$ cubical cells.  All our modelled
haloes will  be spherical, with  a given dark matter  density profile,
mass, and radius.  From now on, these artificial or modelled haloes will
be called template haloes.

Depending on the  test we are analysing, we need  to define the number
of template haloes  we want to study, the  number of time outputs
(different redshifts we look at), as well as the total number of dark
matter particles  to be used.  The  total number of  particles must be
conserved during  the whole evolution to  guarantee mass conservation,
and it must be chosen as a compromise between having a good resolution
in mass for  each halo and the computational  cost.  In the particular
implementation  of all  tests  presented in  this section  we
assumed for simplicity's sake, equal  mass dark matter  particles with
masses $m_{p} \simeq 5.0 \times10^{9}\ M_\odot $.

\subsection{Test 1: Looking for single haloes}

The first test presented is designed  to check the ability of the halo
finder  to look for single  haloes and  compute their  main physical
properties at a given redshift: position, mass, and radius. To achieve this  
we generate an  artificial sample of  haloes with
different numbers of dark matter particles, and with positions, virial
radii,  and virial masses  fixed by  hand.  Then  the halo  finder is
applied to  this mock  universe to verify whether  the detected
haloes agree with those previously made by hand.

\begin{table*}
\centering
\caption{Mean features  of the generated haloes at $z=0$ in test $1$. 
Column 1 contains the halo name;
Cols. 2, 3, and 4 stand for the x, y, and z coordinates respectively
of the centre of mass of each halo in units  of $Mpc$;
Col. 5 shows the total  mass within the virial radius  in units of 
$10^{14}\,M_\odot$;
Col. 6  the virial radius in units  of $Mpc$;
Col. 7 the concentration given by the fitting and between 
parenthesis the concentration of the input density profile;
Col. 8 the density profile inner slope ($\alpha$) given by the fitting;
Col. 9 the density profile outer slope ($\beta$) of the fitting;
Col. 10 the number of dark matter particles within each halo;
Col. 11 the AMR level on which the halo is located.}
\label{tab1}
\begin{tabular}{ccccccccccc}
\hline Halo & x & y  & z & $M_{vir}$ & $r_{vir}$ & {\it C}  & $\alpha$
& $\beta$ & $N_{DM}$ & $level$ \\   
&   (Mpc)   &   (Mpc) &   (Mpc)   & $(10^{14}\,M_\odot)$ & (Mpc) & & & &  \\ 
\hline 
H1 & 0.0 & 0.0 & 0.0 & 20.8 & 2.07 & 6.12 (6.25) & 0.98 & 2.06 & 400000 & 0 \\ 
H2 & -10.0 & -10.0 & -10.0 & 5.19 & 1.30 & 6.92 (7.01) & 0.97 & 2.04 & 100000 & 0 \\ 
H3 & 25.0 & 25.0 & 25.0 & 1.29 & 0.82 & 7.88 (7.87) & 1.006 & 1.99 & 25000 & 1 \\ 
H4 & 10.0 & 10.0 & 10.0 & 0.78 & 0.69 & 8.36 (8.22) & 0.992 & 1.97 & 15000 & 2 \\ 
\hline
\end{tabular}
\end{table*}

Let  us describe  the method  to generate  these artificial haloes.   
Assuming some
general  features  (cosmological  parameters,  number of  time  steps,
number  of  desired  haloes   and  total  number  of  particles),  the
properties of the haloes that populate each time step are made by hand:
the number of  dark matter  particles within each  halo (and  hence, their
masses) and the coordinates of their centres.  With this information
and the cosmological parameters  the average density corresponding to
a  given epoch as  well as  the virial  radius from  Eq. (2)  of the
haloes are computed.  Once the  main physical properties of the haloes
have been  defined, each halo is  created by a  random distribution of
dark  matter   particles  --  using  the   rejection  sampling  method
(\cite{neumann})  -- ,  in  a  way  that its  density profile  is
consistent with a NFW  profile (\cite{nfw97}).

In  this subsection  we  present  a case  characterized  by the  usual
cosmological parameters and with $\sim10^5$ dark matter particles in a
unique  time  step corresponding  to  $z=0$.  For  this test  we  
generated four dark matter haloes with density profiles compatible 
with NFW, in the way explained before. The main   properties   of   
these   mock   haloes   are   summarized   in Table~\ref{tab1}.

\begin{figure*}
\centering\includegraphics[width=12cm]{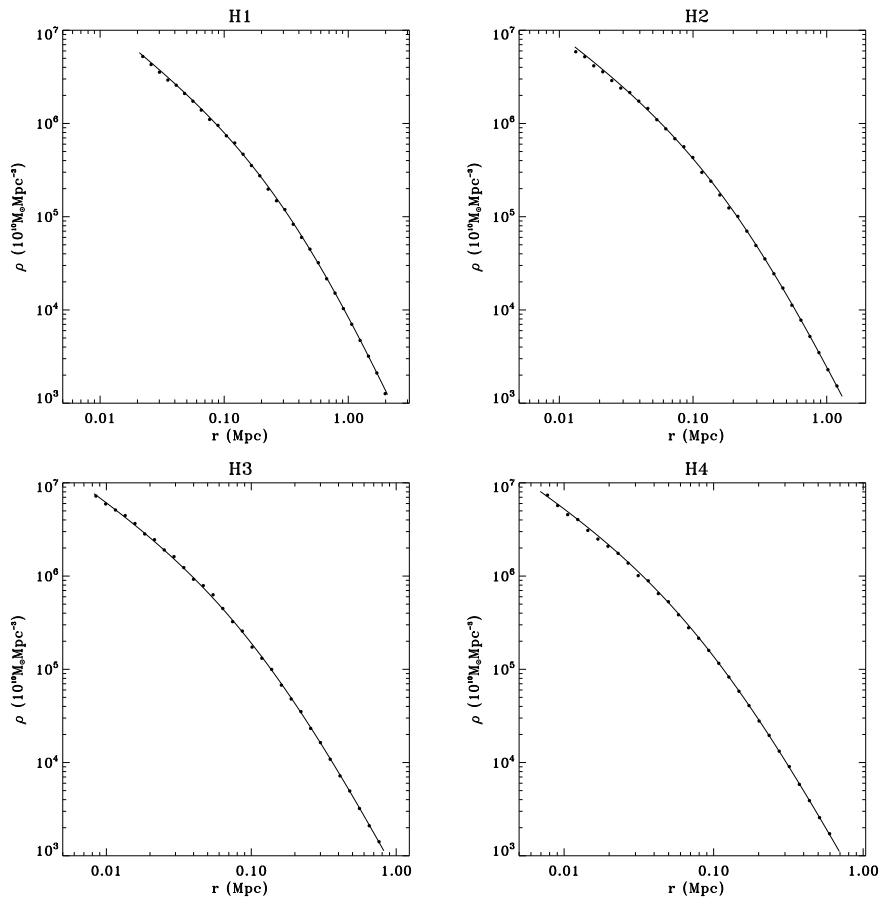}
\caption[Density profiles of the  haloes in test 1]{Radial dark matter
density   profiles  for   the  four   generated  haloes   compiled  in
Table~\ref{tab1}  as a  function  of  the radius  at  $z=0$ (test  1).
Continuous lines stand  for the input profiles of  the mock haloes,
whilst points represent the profiles obtained by the halo finder.}
\label{profiles}
\end{figure*}

The ASOHF halo  finder was applied to  this mock simulation.  The
mean  relative errors  found in  the computation  of the  positions and
radii  are  of the  order  of  $0.1\%$.  The masses  are  perfectly
recovered because  all particles forming  the halo are identified.  Note
that although the results seem excellent, they correspond to
an extremely idealised test.

In Fig.~\ref{profiles}  we plot the radial density  profiles used as
input  to  generate  --  using  the rejection  method  --  the  haloes
(continuous  line)  and  the  obtained  profiles  (dots).   These  last
profiles were computed by  averaging the dark matter  density in
spherical shells of a fixed logarithmic width.

We  fitted  a NFW  density profile  to each  one of  the obtained
profiles.   The concentration,  inner and  outer slopes  are  shown in
Table~\ref{tab1}, respectively. In Col.  7 of this Table, we present
the  concentration  obtained  from   the  fitting  together  with  the
concentration  of   the  density   profile  used  as   input  (between
parenthesis). For the  inner (outer) slope  of the density
profile that we  denoted by $\alpha$ ($\beta$),  the fitted value
must  be compared  with  1.0  (2.0), which  corresponds  to the  value
adopted in the input profile.

We checked that the  errors encountered for the  fitted profiles
are mostly caused  by the rejection sampling method.  In this line we
tested  that the sampling of  an input density  profile with the
rejection  method  produces  particle  distributions that  trace  the
underlying   density  profile   with  a   precision  of   a   few  per
cent. Therefore,  when the  halo finder finds  a halo and  obtains its
density profile,  there is also a  small error when  compared with the
input density  profile.  But  it must be  kept in mind  that this
error does  not arise from  the halo finder  algorithm itself but  
from  the way that this particular test has been set up.

Although this test is very simple, because it only considers four haloes
in a single  time step, it provides us with a  powerful tool to verify
the behaviour of our finder in a very basic situation.  We checked
many  other  configurations (some  of  them  really unrealistic)  with
similar results.   Due to  its clarity and  simplicity we  have chosen
this one to demonstrate our objective.

\subsection{Test 2: Looking for subhaloes}

\begin{figure*}
\centering\includegraphics[width=16cm]{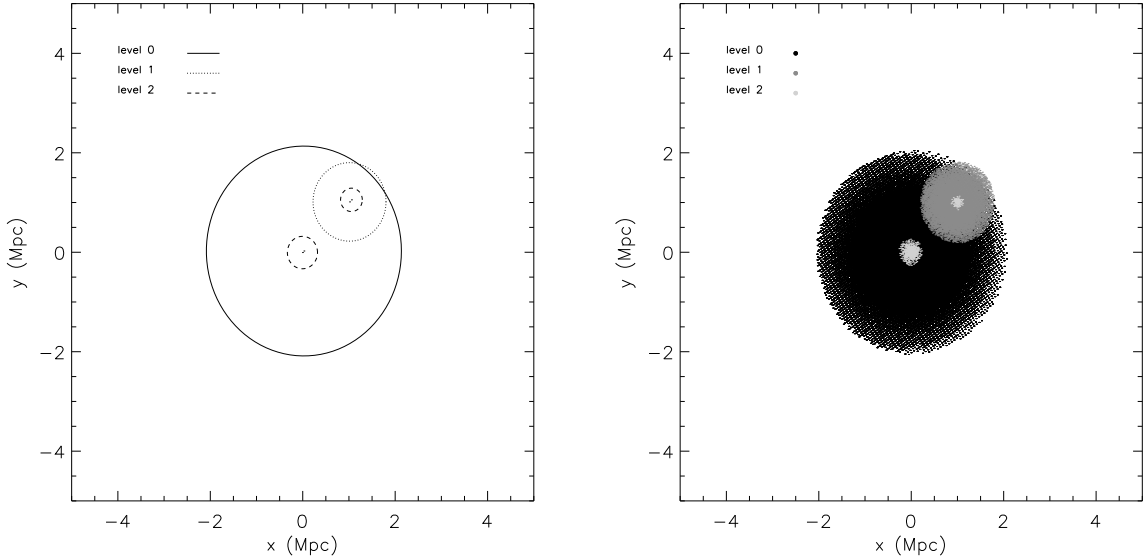}\\
\caption[Substructures in a test halo.]{Mock distribution of a halo
with three substructures: two subhaloes and one sub-subhalo (test 2). 
The left panel shows the 2D projection of the structures found by ASOHF. 
Circles represent the radii of the different structures, 
while the different line types stand for the different refinement levels 
in which the haloes have been found.
The right panel shows the known distribution 
of dark matter particles in this test. 
Different colours stand for the level to which the particles belong.}
\label{test2}  
\end{figure*}

In this  section we present  a simple test  that was  designed to
check the ability of ASOHF to deal with substructures.

Following  the idea  of test  1, we  generated  by hand  a simple
distribution of haloes  placed in different levels of  resolution of a
very basic AMR grid similar to that of the MASCLET code.  For the sake
of simplicity,  only three levels  of refinement (the ground  grid and
two upper  levels) were  considered. The hierarchy  of structures
and substructures generated for this test are distributed according to
these levels.

Among  all configurations that  we  tried for  this test,  we
chose because of its clarity a  simple one in which four structures are
considered: a big dark matter halo in the coarse level hosting two
subhaloes  where at  the  same time  one  of these  subhaloes hosts  a
smaller subhalo, which is a sub-subhalo of the big one.

Figure  \ref{test2} shows  the  configuration analysed  in this  test.
A visual  inspection of  this plot  shows that  the halo
finder also works properly  when dealing with substructures located in
different levels of an AMR grid.

Additionally, the mean relative errors given by ASOHF 
in the estimation of the main properties of the generated
structures are of the order of  
$3\%$, $0.1\%$ and $1\%$ for the mass, 
position, and radius, respectively.  
This value together with  a visual inspection of
Fig. \ref{test2} is an excellent indicator of the good performance of
ASOHF when working with structures that contain different substructures,
at least in a simple configuration like the one considered here.

Because this  configuration is  very basic, we  check this
situation  in  Sect. 5 for a  proper  cosmological simulation  and compare  the
results obtained by ASOHF with those obtained by other well known halo
finders.

\subsection{Test 3: Testing the merger tree}

Once  we checked that the  ASOHF  finder works  properly when it looks  for
single  haloes and  subhaloes, we  checked how well  it computes  the
merger tree for each.

In this  section we consider several  configurations characterized by
the same parameters  as in the previous ones, but  with more than one
time step.  Now the idea is  to generate a given number of haloes, in
the simple way explained before, but forcing different time evolutions
of these haloes.

We are interested in studying  the most common events in the evolution
of dark matter haloes: i) relaxed or isolated evolution, i.e., without
important  interactions or  mergers with  other haloes  (case  I), ii)
ruptures or disruptions of a single halo into two or more smaller haloes
owing mainly to  interactions with the environment or  with other haloes
(case II), and iii) mergers between two or more haloes (case III).  
To do this we chose  four haloes at a given redshift which are
those compiled  in Table~\ref{tab1}, and studied their  evolution in
the three different cases that have into account in a simple way the
most common events explained before.   Then, ASOHF is applied to these
artificial evolutions  to compare the  obtained merger trees  with the
generated ones.

Again, for  the sake  of simplicity and  brevity, a reduced  number of
haloes will be  considered, but note that more
complicated  configurations  were  studied  and  can  be  easily
implemented.

\begin{figure} 
\begin{center}
{\includegraphics[height=14cm]{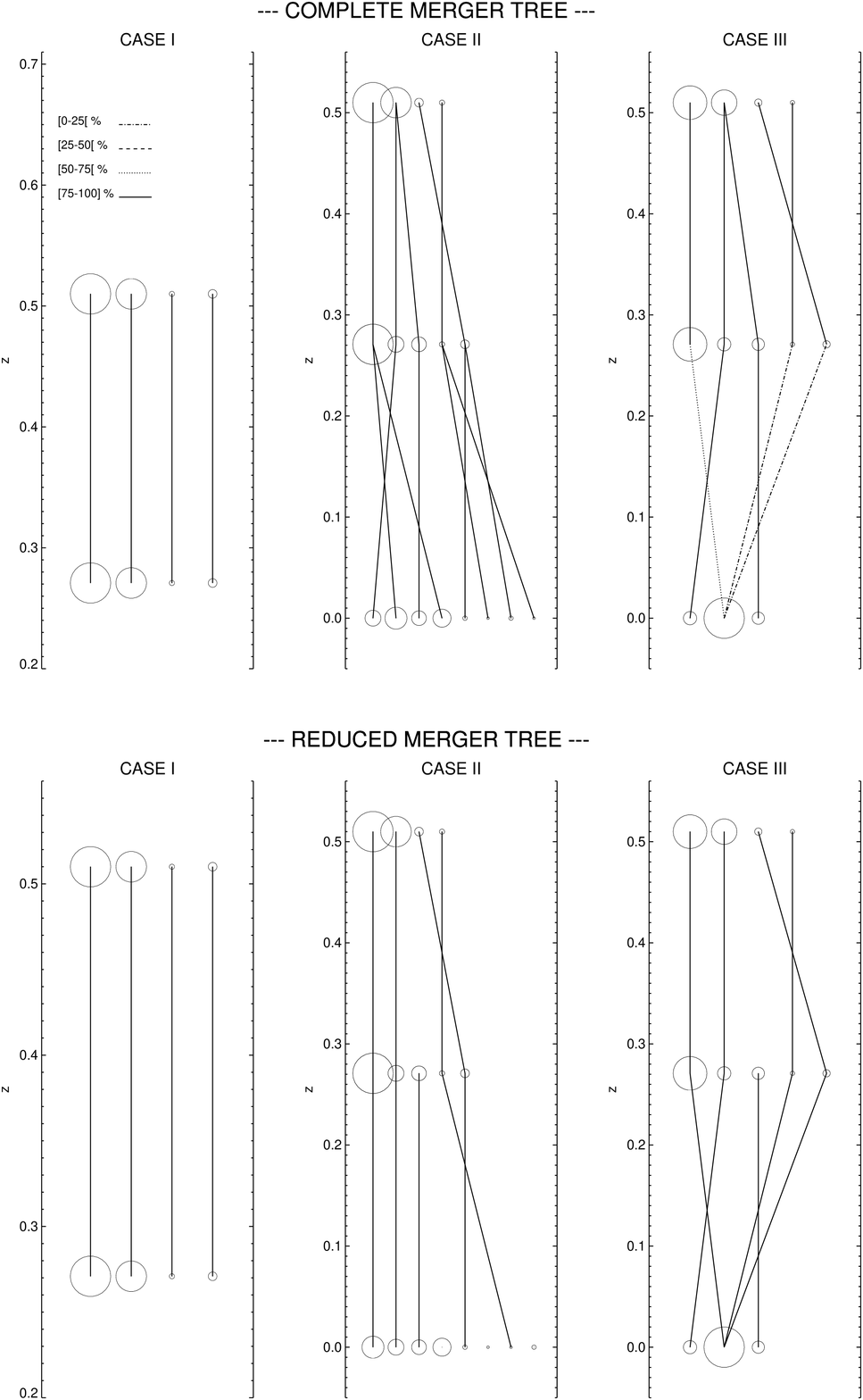}}\\
\end{center}
\caption[Merger trees  of generated haloes.]{Merger  trees for several
haloes in  the three different  cases analysed in Sect.  $4.3$ (test
3). Left, central, and right panels stand for case  I, II, and III,
respectively.  The top panels represent  the results  obtained following
all dark matter particles within each halo (complete merger tree),
whereas the bottom panels stand for the results obtained following only the
closest  particle   to  the  centre  of  each   halo  (reduced  merger
tree). Haloes are represented by circles whose sizes are normalized to
the final mass  at $z=0$.  The different line  types connecting haloes
at  different times in  the upper  plots indicate  the amount  of mass
transferred from the progenitors to their descendants.}
\label{merger_tree_test3} 
\end{figure}

Figure  \ref{merger_tree_test3}  shows the  merger  trees obtained  by
ASOHF in  the different cases  considered here.   The top
panel of Fig. \ref{merger_tree_test3} shows the complete merger tree
obtained for  the four haloes studied  in each of  the three cases
presented here. In the  bottom panel of the same figure, the
same cases  are represented but following only the  
closest particle to  the centre of each halo (reduced merger tree).

The    line     segments    joining    the     circles    in    Fig.
\ref{merger_tree_test3}  are a  relevant feature  of the  plot because
they indicate that the halo at the earlier time is considered to merge
into (or to be identical to)  the halo at the later time. Moreover, in
the upper panel the different  line types represent the percentage of
mass that  goes from one halo to  another. Thus, a halo  at later time
connected with a halo at earlier time by a dash-dotted line means that
up to  25\% of its  total mass comes  from that halo at  earlier time.
The same idea applies to the other line types.

The horizontal  axis is designed  to separate the haloes  according to
their future  merging activity.  It  does not directly  indicate space
positions, although there should be some correlation between how close
two haloes  are in  the plot and  how close  they are in  space (because
haloes  need  to  be close  to  merge later  on).
The vertical axis shows the redshift  of each time step in the simulation.
The  size  of each  circle  indicates the  virial  mass  of each  halo
normalized to its  final mass at the last  iteration. Because the iterations
go in descending  order, the last iteration corresponds  to the lowest
redshift.

The different cases analysed  in this section and their representation
in  Fig.  \ref{merger_tree_test3} are  discussed  in  detail in  the
subsections below.

\subsubsection{Case I}  

In this first case, we study the most trivial situation. Only two time
steps  corresponding  to   $z  \sim  0.5$  and  $z   \sim  0.3$  are
considered. The objective is that the four haloes generated at $z \sim
0.5$ would  be exactly the  same as those at $z \sim  0.3$.  The
selection of the redshifts is made in order to obtain a fast evolution
of the haloes, but  it is possible to make it for  time steps much more
realistic and separated  in time.  In any case,  this selection is not
relevant to achieve our objective, which is to check the performance of the
halo finder computing the merger tree of a given halo, i.e., following
the  particles that  belong to  a halo  at a  given epoch  through the
evolution.

The haloes at  different epochs are generated in  the way explained 
above.   In this particular  case, we force that  the four
considered haloes  at both epochs  would be identical, i.e.,  with the
same  particles in  each one  and  with the  same radius  and
position of the centre.

In  order to  be as  clear as  possible when  talking about  haloes at
different epochs, we will use the notation $h_{ij}$, where $i$
stands for the iteration  number (iterations in descendant order
and then corresponding the iteration or time step $1$ to the lowest
redshift)  and   $j$  for  the   halo  number  in  the   iteration  i,
respectively.

According to this notation, the generated relations between the haloes
in this case are

\begin{itemize}
\item [\textbullet] $h_{21} \Longrightarrow h_{11} [100\%]$ 
\item [\textbullet] $h_{22} \Longrightarrow h_{12} [100\%]$
\item [\textbullet] $ h_{23}  \Longrightarrow h_{13} [100\%]$
\item [\textbullet] $h_{24} \Longrightarrow h_{14} [100\%]$ .
\end{itemize}
 
These connections tell  us we are working only with  two time steps (2
corresponding to $z \sim 0.5$ and 1 to $z \sim 0.3$, respectively) and
each one of these  epochs has four haloes (j runs from  1 to 4 in both
time steps).  In addition,  the number between square brackets informs
us about the percentage of the  mass of each halo that it obtains from
its progenitor.  For instance, the  first relation tells us that the $100
\%$  of  the  mass of  the  halo  number  one  in the  last  iteration
($h_{11}$) comes  totally ($100\%$)  from the halo  number one  in the
previous time step ($h_{21}$).

We applied the halo finder  to this artificial  evolution and
constructed the  merger tree of the selected  haloes tracking all
dark  matter particles  for a given  halo backwards in  time.  For
each halo in this case, two merger trees, the complete and the reduced
one,    have    been   built.     The    left    panels   of    Fig.
\ref{merger_tree_test3} show the obtained results.  The
upper-left plot of this figure  shows the complete merger tree for the
considered haloes, whereas the lower-left plot represents
the reduced merger tree for the same haloes.

According  to these  plots it  is evident  that for  case I  the halo
finder tracks the correct history for all considered haloes.

Additionally,  the connection  lines linking  the haloes  at different
redshifts in the  complete merger tree inform us  about the percentage
of mass  that each  younger halo receives  from its  progenitors. In
this particular case,  this percentage is in all  cases greater than
$75\%$,  in perfect  agreement  with the  expected results  ($100\%$).
More precisely, the obtained results are

\begin{itemize}
\item [\textbullet] $h_{21} \Longrightarrow h_{11} [99.97\%]$ 
\item [\textbullet] $h_{22} \Longrightarrow h_{12} [99.98\%]$
\item [\textbullet] $h_{23}  \Longrightarrow h_{13} [99.99\%]$
\item [\textbullet] $h_{24} \Longrightarrow  h_{14} [99.99\%]$ .
\end{itemize}

Although in  this example the  obtained percentages are  very accurate
with regard  to the expected ones,  we should point out  that they are
not always exact.  This is because although the particles
are  forced to  be  the same  and belong  to  a given  halo, they  are
distributed randomly.  Thus, 
the  haloes are not exactly found within the same boundaries by the finder
for  different redshifts in this test 
and then some particles are not taken into account.

\subsubsection{Case II}

In this  case we  are interested in  checking the capabilities  of the
halo finder when some haloes  suffer one or several disruptions during
their  evolution, when  they lose  mass and  reduce their  size.  This
process operates at  two regimes for different reasons.   This is quite
common  in very  small size  haloes. The  reason is  that these
haloes  are not  really  gravitationally well   bound and  can
easily  be disrupted  by  interactions  with environment  or  with  other
haloes.  For larger haloes, those mass losses are smaller and they are
usually associated with tidal interactions.

To  study this case  we started  at $z  \sim 0.5$  with the  four haloes
summarized in  Table~\ref{tab1}.  Now, three time steps  of the haloes
evolution corresponding  to $z \sim 0.5$,  $z \sim 0.3$  and $z=0$ are
considered.  The  results of  this situation are  shown in  the middle
panels (top and bottom) of Figure \ref{merger_tree_test3}.

Here the  generated  evolutions can  be  summarized in  the
following  relations.  In  a first  step, the  connections  between the
haloes of the third ($z \sim 0.5$) and the second ($z \sim 0.3$) iterations 
are

\begin{itemize}
\item [\textbullet] $h_{31} \Longrightarrow h_{21} [100\%]$ 
\item [\textbullet] $h_{32} \Longrightarrow h_{22} [100\%] , 
h_{23} [100\%]$
\item [\textbullet] $h_{33} \Longrightarrow h_{24} [100\%]$ 
\item [\textbullet] $h_{34} \Longrightarrow h_{25} [100\%]$ .
\end{itemize}

On  the other  hand, the  links between  the haloes  generated  in the
second iteration ($z \sim 0.3$) with those in the first one ($z=0$)
are

\begin{itemize}
\item [\textbullet] $h_{21} \Longrightarrow h_{12} [100\%] , 
h_{14} [100\%]$
\item [\textbullet] $h_{22} \Longrightarrow h_{11} [100\%]$ 
\item [\textbullet] $h_{23} \Longrightarrow   h_{13} [100\%]$  
\item [\textbullet] $h_{24} \Longrightarrow  h_{16} [100\%] ,
 h_{18} [100\%]$ 
\item [\textbullet] $h_{25} \Longrightarrow  h_{15} [100\%] , 
h_{17} [100\%]$ .  
\end{itemize}

As a  result  of the  whole evolution  there are  eight
haloes instead  of the first four,  which were generated in the
same manner as  in the previous case, but which  were forced to
share a certain  number of particles with their  ancestors, which
property is the key to build their evolution history.

After building these artificial  evolutions, ASOHF was applied to
this mock universe to obtain the merger trees of the involved
haloes.

As  we  can  see  in   the  upper-middle  plot  (case  II)  of  Fig.
\ref{merger_tree_test3}, the halo  finder again provides  very accurate
results,  in perfect  agreement  with those  exposed  before.  In  all
cases  the obtained  percentages  are between  $99.9\%$ and  $100\%$.
Again, the value  of the percentages can be explained  if we take into
account  that each  halo has  been populated  with  particles randomly
placed.  Then,  the particles  positions are not  always the  same and
small deviations are expected.

If we compare the  upper-middle plot of Fig. \ref{merger_tree_test3}
(complete  merger tree)  with  the lower-middle  plot (reduced  merger
tree), the results  completely agree but  in the lower plot
each halo is only allowed to have  one descendant at maximum.\\

\subsubsection{Case III} 

Here the response of the halo finder in a merger
between two or more haloes is checked.

To analyse  this, we  started again with  the same haloes  and time
steps as before.  Now, the different evolutions can be
summarized with the following links.  In a first step the connections
between the  haloes of the third ($z \sim 0.5$)  and the second 
($z \sim 0.3$) iterations are (the same as in the previous case)

\begin{itemize}
\item [\textbullet] $h_{31} \Longrightarrow h_{21} [100\%]$ 
\item [\textbullet] $h_{32} \Longrightarrow h_{22} [100\%] , 
h_{23} [100\%]$ 
\item [\textbullet] $h_{33} \Longrightarrow h_{24} [100\%]$ 
\item [\textbullet] $h_{34} \Longrightarrow  h_{25} [100\%]$ .
\end{itemize}

But now the links between the haloes of the second ($z \sim 0.3$) and 
first ($z=0$) iterations are 

\begin{itemize}
\item [\textbullet] $h_{21}+h_{24}+h_{25} \Longrightarrow h_{12}
 [74.49\% + 9.48\% + 16.03\%]$
\item [\textbullet] $h_{22}\Longrightarrow h_{11} [100\%]$ 
\item [\textbullet] $h_{23} \Longrightarrow h_{13} [100\%]$ .
\end{itemize}

In the  end three haloes  are obtained as  a result of  the different
processes that happened during their evolution.

From the right panels of Fig.~\ref{merger_tree_test3} (top and bottom
plots), we can  deduce that despite the triple  merger that has taken
place in  the last time step,  the halo finder  provides very accurate
results.  Indeed, the  results obtained for this merger  event are

\begin{itemize}
\item [\textbullet] $h_{21}+h_{24}+h_{25} \Longrightarrow h_{12} 
[74.45\% + 9.45\% + 16.08\%]$ ,
\end{itemize} 
where the percentages perfectly agree with the expected ones.

\subsection{Test 4: Analysing a sample of haloes}

\begin{figure}
\begin{center}
\scalebox{0.1}
\centering\includegraphics[height=10cm] {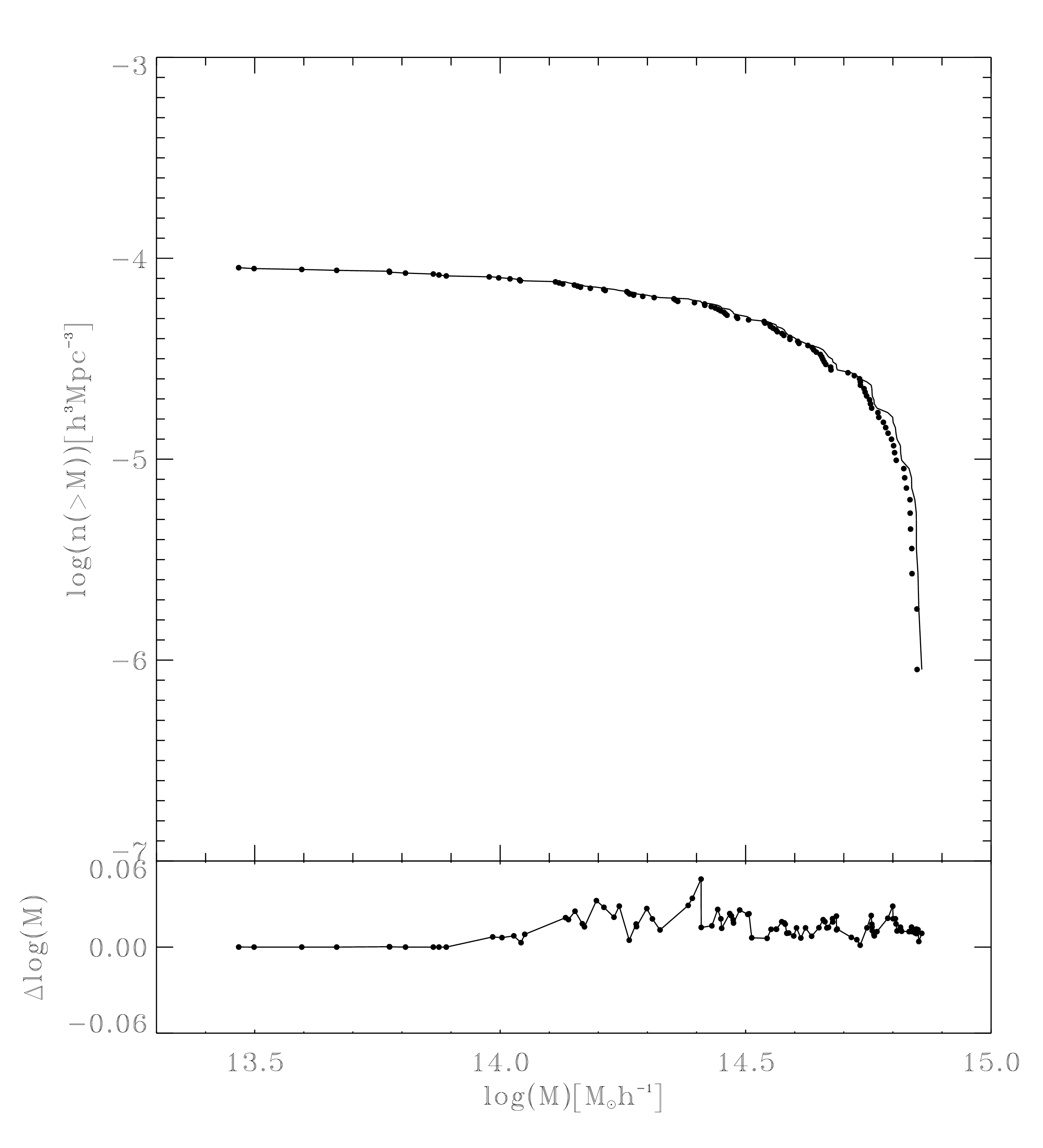}
\end{center}
\caption[Artificial mass function.]{Top panel: academic mass function 
corresponding to the generated sample of $100$ haloes for test 4. 
Dots represent the mass function obtained by ASOHF, whereas the
continuous line corresponds to the function generated by hand.
Bottom panel: relative error or difference in mass between the two 
distributions shown in the upper plot.}
\label{mass_function_test4}
\end{figure}

The  analysis of  big samples  of  haloes is  crucial in  cosmological
applications. Therefore,  once we checked the halo  finder works
correctly  looking for  single  haloes and  constructing their  merger
trees, we should check what its response is when working with a large
sample of haloes and computing all their properties.  Once this sample
of  haloes  is  built,  an  academic mass  function,  i.e.,  the  mass
distribution of all the generated haloes, is computed.

For  the sake  of simplicity,  only  one time  step corresponding  to
$z=0$  was  considered.  Then  a  sample of  $100$ haloes  with
masses randomly distributed  between $1.0 \times10^{13}\, M_\odot$ and
$1.0\times10^{15}\, M_\odot$, was  generated. The position of the
centre and the radius of  each halo are obtained randomly, whereas the
number  of particles belonging  to each  one of  them is  derived from
their masses.

Once this  mock universe  was generated, the  ASOHF finder  was applied.
Then the  academic mass  function of the  well known  distribution of
haloes  is compared with  the mass  function of  the sample  of haloes
obtained by ASOHF.

The   results   obtained   in   this   case  are   shown   in   Fig.
\ref{mass_function_test4}, in  which the number  of objects of  a given
mass  is plotted  as a  function  of the  mass.  This  plot shows  the
theoretical or  academic mass function  (continuous line) and  the one
obtained  by  ASOHF   (filled  dots).   As  we  can   see,  these  two
distributions  are almost completely  superposed.  As  a proof  of the
precision of the finder we can compare the masses of the most and less
massive haloes of  the sample obtained by the  two methods.  Thus, the
most  massive  halo  found  by  the  finder  has  a  mass  of  $9.8622
\times10^{14}\,  M_\odot$, whereas this  halo was  supposed to  have a
mass of $9.8623 \times10^{14}\, M_\odot$. The same occurs for the less
massive halo, which  was found by the finder to have  a mass of $1.48
\times10^{13}\, M_\odot$,  whereas it was  supposed to have a  mass of
$1.49 \times10^{13}\,  M_\odot$.  In  addition, as we  can see  in the
bottom panel of Fig.  \ref{mass_function_test4}, the maximum value of
the relative  error in mass  between the theoretical and  the obtained
mass functions is $\sim 5\%$.

\section{Comparison with other halo finders}

In this section we compare the results of ASOHF with two other halo-finding   
mechanisms,  namely   AFoF   (\cite{vanKampen95})  and   AHF
(\cite{Knollmann}).   We applied  these three  halo finders  to a
cosmological  simulation  carried   out  with  the  cosmological  code
MASCLET. The main  properties of this simulation are  explained below.

For  the AFoF  run,  a linking  length  of $0.16$  times  the mean  DM
particle  separation was used,  yielding an  overdensity  at the
outer radius  comparable to the  virial overdensity used in  the ASOHF
run.   This  linking length  is  obtained  when  scaling the  standard
linking length  of $0.2$  by $(\Delta_c/ \Omega)^{-1/3}$  according to
the adopted cosmology (\cite{eke96}).

For the run with  AHF, we used a value of  $5$ for the parameters
with regard to the  refinement criterion on the domain  grid (DomRef) and
on the refined grid (RefRef), respectively.  To understand the role of
these parameters we need to explain briefly how AHF operates.  
Once the user  has provided the particle distribution,  the first step in AHF
consists in covering the whole simulation box with a regular grid of a
user-supplied size.   In each cell the particle  density is calculated
by  means  of  a   triangular  shaped  cloud  (TSC)  weighting  scheme
(\cite{hockney88}). If the particle  density exceeds a given threshold
(the refinement  criterion on  the domain grid,  DomRef), the  cell is
refined and covered with a finer  grid with half the cell size. On the
finer grid (where it exists),  the particle density is recalculated in
every cell and  then each cell exceeding another  given threshold (the
refinement criterion on refined  grids, RefRef) is refined again. This
is repeated until a grid is  reached on which no further cell needs to
be  refined.  Following  this  procedure yields  a grid  hierarchy
constructed in  a way  that it traces  the density field  and can
then be  used to find  haloes and subhaloes  in a similar way  to that
used by ASOHF.

In all the runs, an equal  minimum number of dark matter particles per
halo was considered.  This  number has been set to $50$ particles
per halo.  In spite of  this consideration, we expect some differences
in the  final results obtained  with the different halo  finders.  The
main explanation for  these expected discrepancies has to  do with the
different techniques used  by the three methods in  the generation of
the  density  field  and  hence  in the  definition  of  the  haloes.
However, general properties of  the simulation should be well described
by the three finders.

\subsection{Simulation details}

The simulation described  here was  performed with the
cosmological  code MASCLET  (\cite{quilis04}).  This  code  couples an
Eulerian  approach  based  on  {\it high-resolution  shock  capturing}
techniques  for describing  the  gaseous component  with a  multigrid
particle mesh  N-body scheme for evolving  the collisionless component
(dark matter).  Gas and dark matter are coupled by the gravity solver.
Both schemes benefit  by using an AMR strategy,  which permits them to gain
spatial and temporal resolution.

The numerical simulation was performed assuming a spatially flat
$\Lambda CDM$  cosmology with the  following cosmological parameters:
matter  density  parameter,  $\Omega_m=0.25$;  cosmological  constant,
$\Omega_{\Lambda}=\Lambda/{3H_o^2}=0.75$;  baryon  density  parameter,
$\Omega_b=0.045$;  reduced Hubble  constant, $h=H_o/100  km\, s^{-1}\,
Mpc^{-1}=0.73$;  power  spectrum index,  $n_s=1$;  and power  spectrum
normalization, $\sigma_8=0.8$.  The initial  conditions were set up at
$z=50$, using a  CDM transfer function from \cite{EiHu98}  for a cube
of  a comoving side  length  $ 47\,  h^{-1}\,  Mpc$.  The  computational
domain was discretised with $256^3$ cubical cells.

This  simulation uses  a maximum  of six  levels of  refinement, which
gives  a peak spatial  resolution of  $3\,h^{-1}\,kpc$.  For  the dark
matter two particles species were considered to be the best mass
resolution  $\sim  4\times  10^7\,  h^{-1}\, M_\odot$,  equivalent  to
distribute $256^3$ particles in the whole box.

\begin{table}
\centering 
\caption{General  results obtained by  ASOHF, AFoF  and AHF  at $z=0$.
Column  2  stands   for  the  number  of  detected   haloes  with  masses
$\geq1.0\times10^{12}\,M_\odot  h^{-1}$,  whereas   Cols.  3  and  4
represent the  minimum and maximum masses  of all the  found haloes in
units of $10^{9}$ and $10^{14}\,M_\odot h^{-1}$, respectively.}
\label{tab4}
\begin{tabular}{cccc}
\hline Halo Finder & $N_{haloes}$ & $M_{min}$  & $M_{max}$  \\   
&$(M>10^{12}\,M_\odot h^{-1})$ & $(10^{9}\,M_\odot h^{-1})$   & 
$(10^{14}\,M_\odot h^{-1})$ \\ 
\hline 
ASOHF & 157 & 4.5 & 5.9  \\ 
AFoF & 130 & 7.3 & 6.4  \\ 
AHF & 181 & 4.9 & 5.9  \\ 
\hline
\end{tabular}
\end{table}

\begin{figure}
\begin{center}
\scalebox{1.5}
\centering\includegraphics[height=100mm] {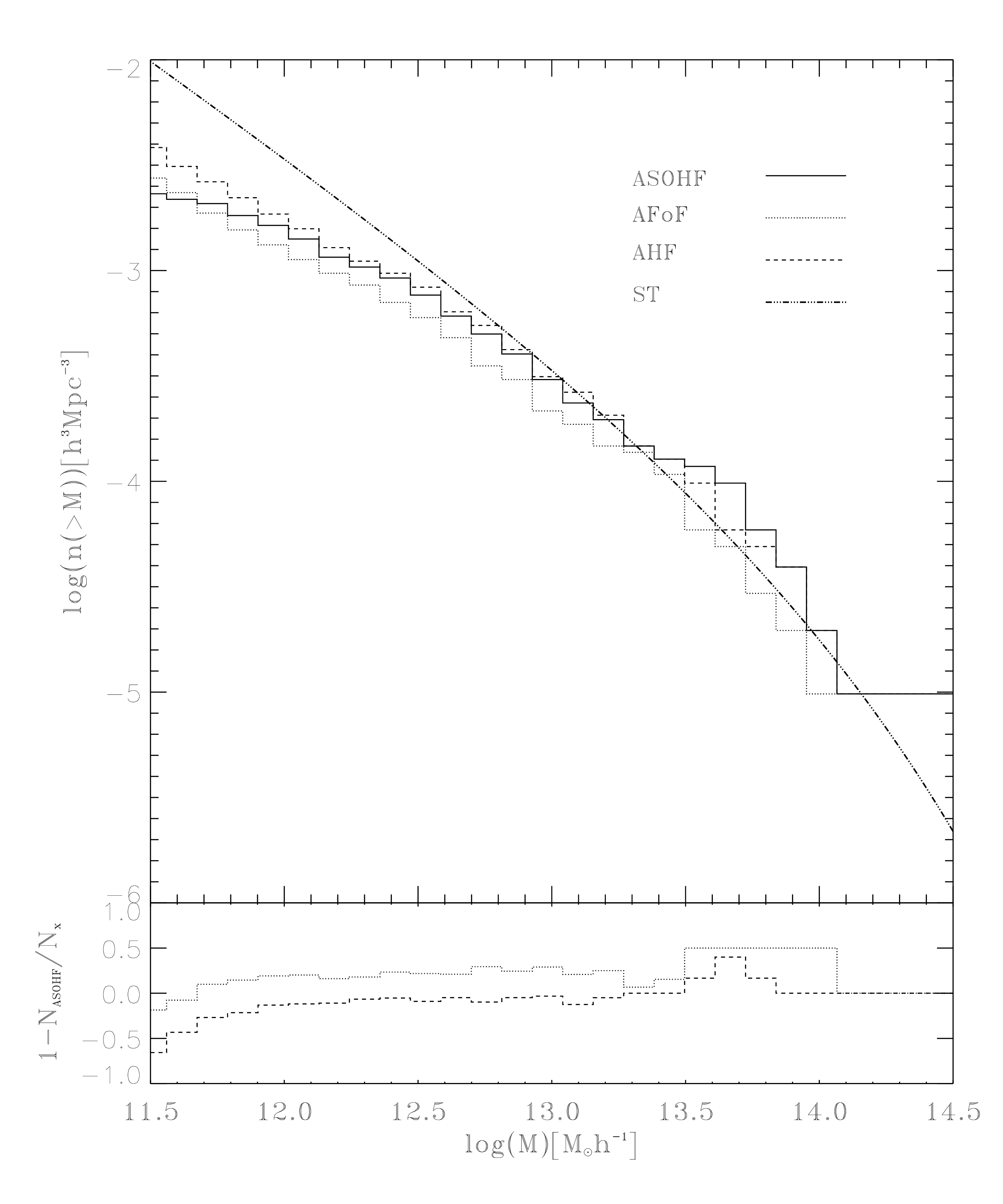}
\end{center}
\caption[Mass function for the sample of haloes at $z=0$]{Top panel:
comparison of  the mass functions obtained  by ASOHF, AHF  and AFoF at
$z=0$.  The mass function predicted  by Sheth \& Tormen is also shown.
Bottom panel: relative difference in  the number of haloes between AFoF
and AHF compared to ASOHF.}
\label{real_mass_function}
\end{figure}

\begin{figure*}
\centering\includegraphics[height=90mm] {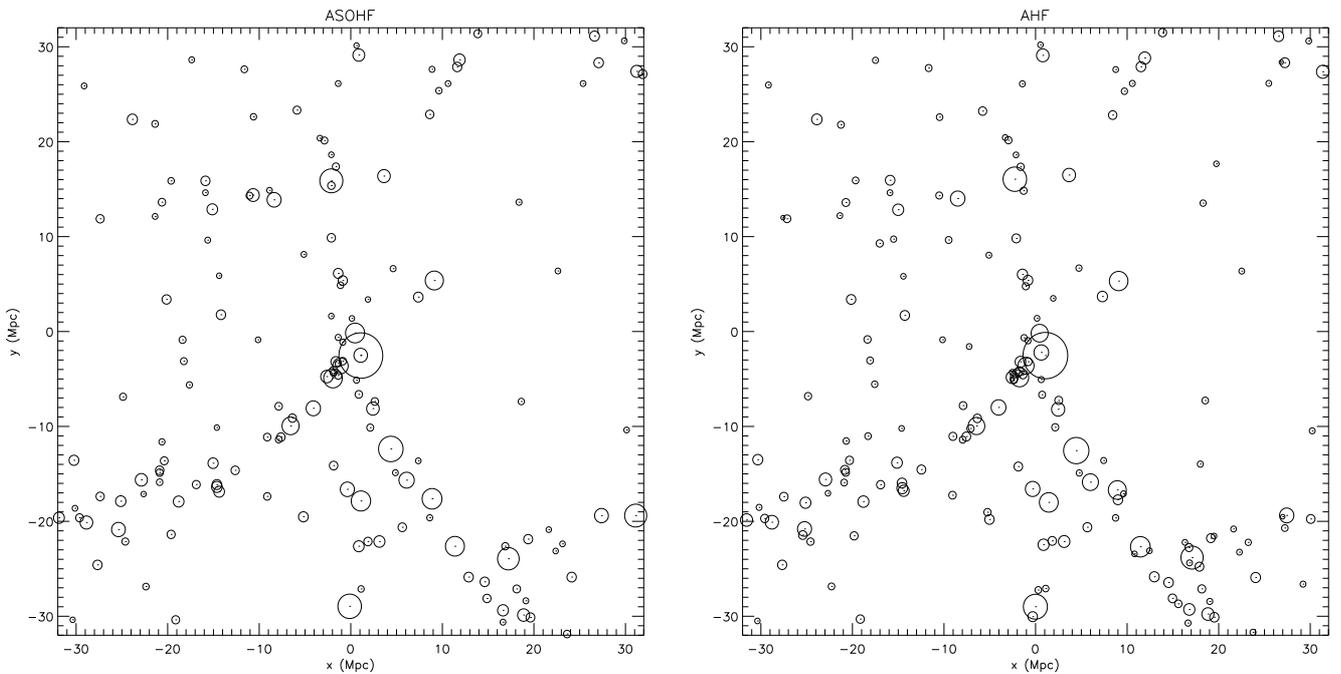}
\caption[2D projection  for haloes at $z=0$]{2D  projection for haloes
at $z=0$. Only haloes with masses above 
$1.0\times10^{12}\,M_\odot h^{-1}$ are shown.
Left panel stands for haloes and subhaloes found by ASOHF,
whereas right panel corresponds to those  found by AHF. The size of the
different haloes is given by their virial radii.}
\label{proyecciones}
\end{figure*}

\subsection{Halo mass function}

Here we  present the  sample of  haloes obtained  from the
cosmological  simulation by  the  three halo  finders used  in the  present
study,  namely  ASOHF,  AFOF   and  AHF.  Their  main  properties  and
differences are discussed.

The three halo finders obtained a relatively large sample of galaxy
clusters and groups spanning an approximated range of masses from $1.0
\times10^{9}\,M_\odot  h^{-1}$ to  $2.0\times10^{14}\,M_\odot h^{-1}$.
The total number  of structures identified by ASOHF,  AFoF, and AHF has
been $1339$,  $7448$, and  $1712$, respectively.  Although  the numbers
and masses  of the  detected haloes are  roughly consistent, they  are, as
expected,  slightly different  among them.   
These results  are more
similar  between ASOHF and  AHF, whereas  AFoF identifies  more smaller
haloes.

To analyse the  simulation mass function, we restricted ourselves 
to study the best-resolved haloes, that is those haloes with masses 
above $1.0\times10^{12}\,M_\odot h^{-1}$. 
The  number of haloes with masses  above this limit and
the maximum and minimum masses (in all the sample) of the found haloes
by the different halo finders are summarized in Table~\ref{tab4}.  The
obtained results by the three  finders, although very similar, are not
exactly the same.  This was expected because  each halo finder uses
different  approximations and techniques.  ASOHF
uses  the  grid hierarchy  generated  by  the cosmological  simulation
itself, whereas  AHF has  to construct  a new set  of grids with different
criteria because  only a list  of particles is  provided to
them which is consequently not identical with that used by ASOHF.

In  Fig.~\ref{real_mass_function} we compare  the mass  functions at
$z=0$ of the  simulation as obtained by the  different halo finders in
this study.   We also  present a  comparison with  the mass
function proposed by \cite{ST} (ST).

\begin{figure*}
\centering\includegraphics[height=150mm] {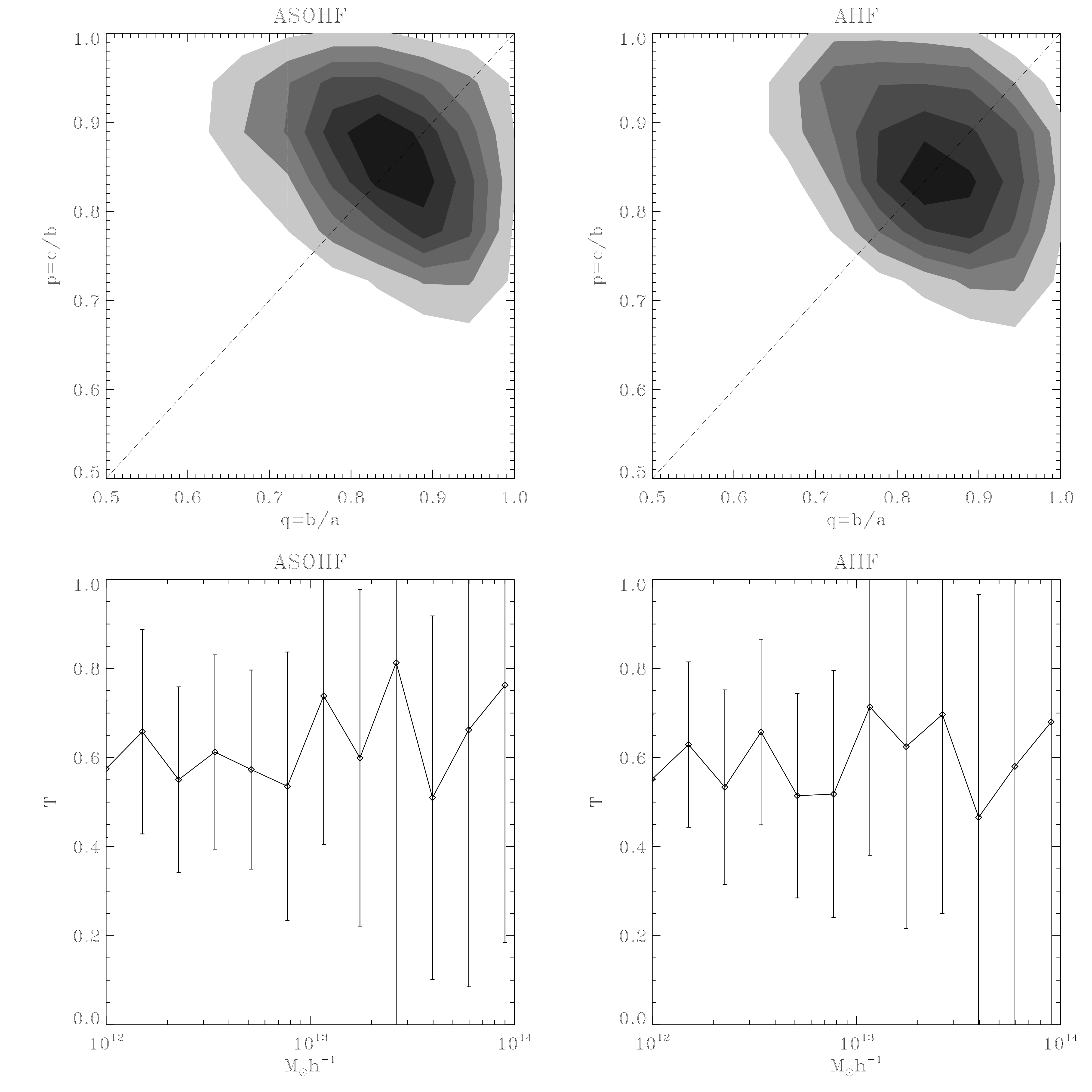}
\caption[Normalized distribution of shapes of haloes.]{Distribution of 
the halo shapes at $z=0$ as found by ASOHF (left panels)
and AHF (right panels), respectively. Only haloes with masses above
$1.0\times10^{12}\,M_\odot h^{-1}$ are taken into account.
The top plots show the distribution of $p=c/b$ against $q=b/a$. 
The dashed line
represents the division between oblate and prolate haloes.
The shaded grid was computed by binning individual haloes to a 
two-dimensional grid. Six contour lines equally spaced are plotted 
to highlight the shape of the two-dimensional distributions.
The bottom panels show the triaxiality parameter as a function of the 
halo masses. The error bars represent $\sqrt N$ uncertainties due 
to the number counts in the different mass bins.}
\label{shapes}
\end{figure*}

The obtained mass functions  show a considerable dispersion mainly in
the lower limit  of mass compared with the  ST prediction.  Note though 
that the theoretical  mass function proposed  by ST has
been    calibrated   using    an    overdensity   of    $\Delta_c=174$
(\cite{tormen98}), whereas in our case this overdensity is $\sim 374$.
The  bottom  panel  of  Fig.~\ref{real_mass_function}  displays  the
relative  deviation of  the mass  functions obtained  by AFoF  and AHF
with respect to the results produced by ASOHF.  Hence, a positive deviation
means that the  ASOHF run found more haloes in the  given bin than the
halo finder it is  compared with.  Generally speaking, we  find good agreement
between  the three  mass  functions, although  ASOHF  and AHF  results
exhibit  a better  resemblance, which is expected  because the
similarities of both methods.

Let  us  point out  that  the dispersion  of  the  mass function  when
compared with the reference mass  function (ST) is a well known issue.
We stress  that is  out of the  scope of this  paper to
discuss how  representative the considered  simulation is. Instead, we
use this simulation to test whether the different halo finders 
produce similar  results. In this  sense, we emphasize that  the three
algorithms compared agree very well.

For the  sake of completeness, we mention that  the dispersion of
the mass function  is a complex topic that is abundantly discussed in the 
literature. A few
examples, among  many others, could  be: i) the work  by \cite{reed07},
where the authors  study the dispersion  of the mass function  for several
simulations depending  on the redshift,  ii) the results of  the GIMIC
project  (\cite{crain09}), where  an important  dispersion in  the mass
function is  shown depending  on the considered  region, and  iii) the
dispersion of the mass function found by \cite{yaryura10} related with
very large structures.

From now  on, we  restrict ourselves to  analyse only the  main differences
between the  AHF and the ASOHF codes, that  is, between the "grid
based on halo finders".  The  reason is that these methods are more
directly comparable with each other.  Still, given that we use AHF as
a stand-alone halo finder, differences are expected.
  
To have a  first order comparison of the spatial distribution
of  the haloes  encountered by  both  codes, Fig.~\ref{proyecciones}
shows the 2D projection along the z axis of the simulated box of all
haloes (and/or subhaloes) found by  ASOHF and by AHF at $z=0$.  We
only  show   those  haloes  or  subhaloes  with   masses  larger  than
$1.0\times10^{12}\,M_\odot   h^{-1}$.     Both   panels   are   highly
consistent.  All  relevant features of the  halo distribution were  
caught with  both methods,  and therefore  they  seem perfectly
comparable.

The  main differences  between both  methods arise  when  the smallest
structures  found  are  taken   into  account.   Whereas  the  biggest
structures are  perfectly recognized in both plots,  the smallest
represent  the   main  source   of  disagreement.   These
discrepancies may have  a variety of causes, of which  the most important
is that the  finders make use  of very different  techniques 
to  compute the dark  matter density distributions.   Both codes
create  their  structures  of  nested  grids  according  to  different
criteria. Therefore, a slight change in the number of grids, especially
for the small  objects, could alter the way in  which these objects are
resolved, making  them detectable or  not.  Leaving this issue aside,
both distributions are fully comparable.

\subsection{Halo shapes}

The shapes of haloes are described by the axes, $a \ge b\ge c$, of the
ellipsoid  derived from the  inertia tensor,  as described  in Sect.
$3.3$.   For   the  sake  of   completeness,  we  have   compared  the
distribution of halo shapes obtained  by ASOHF and AHF for haloes with
masses above $1.0\times10^{12}\,M_\odot h^{-1}$.  The obtained results
are shown in Fig. \ref{shapes}.  
As  we can  deduce from
these  results,  haloes  are  generally  triaxial  but  with  a  large
variation in  shapes. Prolate objects have $p=1$,  oblate objects have
$q=1$, and spherical  objects have $p=q=1$. 

Our results  show that the
haloes  are  mainly  spherical   but  with  a  slight  preference  for
prolateness  over oblateness.  This distribution  qualitatively  agrees
with   previous    results   (e.g.,   \cite{frenk88},   \cite{cole96},
\cite{bailin05}).

\begin{figure}
\centering\includegraphics[height=130mm] {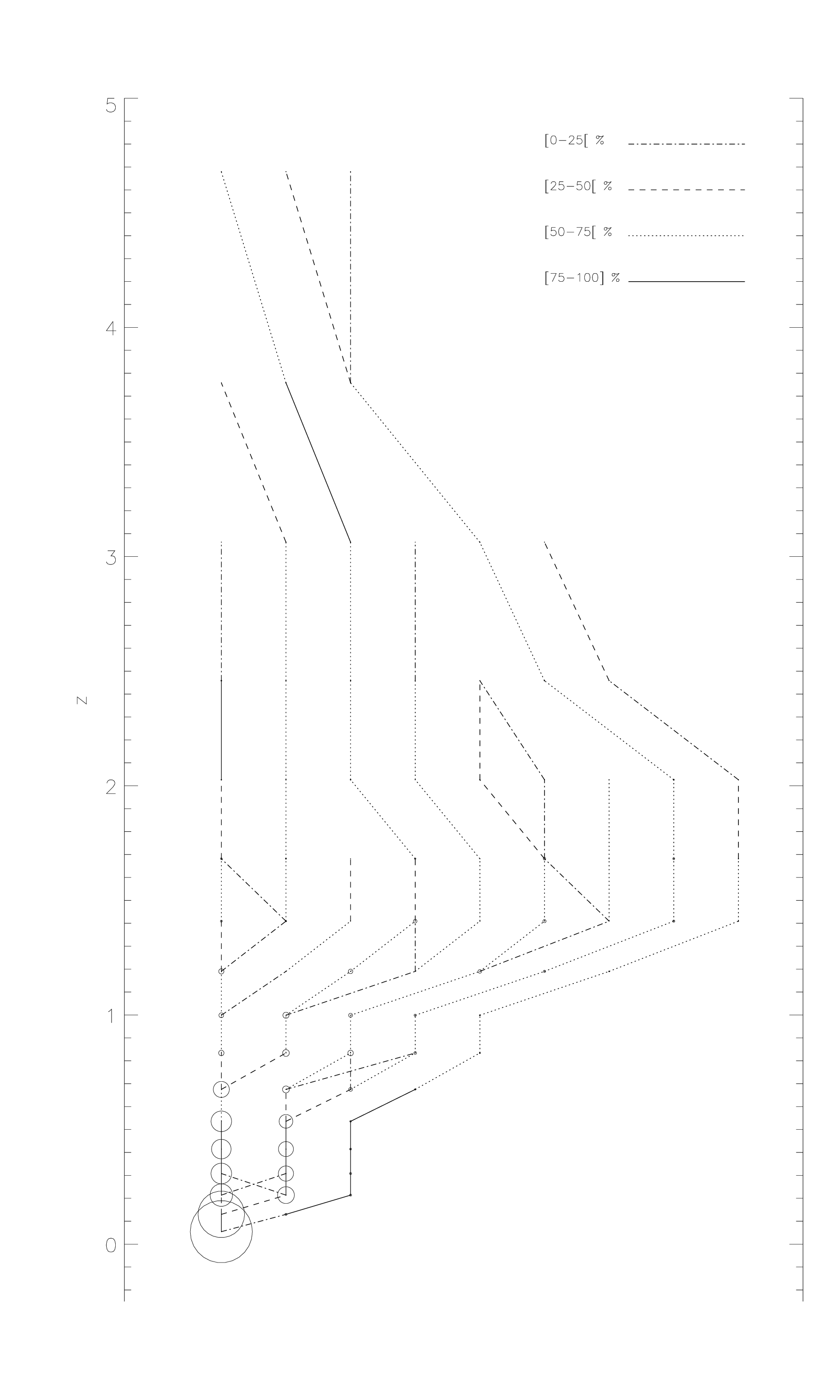}
\caption[Merger tree]{Merger tree of  the analysed host halo.  Cluster
haloes are  represented by circles  whose sizes are normalized  to the
final  mass at  z  = 0.  Lines  connecting haloes  at different  times
indicate the amount of mass  transferred from the progenitors to their
descendants. Only the contribution of the most massive progenitors
is displayed.}
\label{merger_tree}
\end{figure}

Bottom panels of Fig.  \ref{shapes} show the triaxiallity parameter,
T, of the haloes found by ASOHF  (left plot) and AHF (right plot) as a
function of halo masses.  In  this figure the x-axis was divided
into $12$ mass  bins equally spaced in logarithmic  scale, and the error bars
represent  $\sqrt  N$ uncertainties  due  to  the  number counts.  The
general  trends  obtained  from  these  plots agree  with
previous     results     (e.g.,    \cite{warren92},     \cite{shaw06},
\cite{allgood06}).   As it  would  be naively  expected, more  massive
haloes tend to  be less spherical and more  prolate. In a hierarchical
model of structure formation, more massive haloes form later, and have
less  time to  relax and  to  form more  spherical configurations.  In
addition, because  haloes tend to  be formed by matter  collapsing along
filaments,  they   generally  lead  to  prolate   rather  than  oblate
structures.  Because our halo sample is statistically small, the general
trend obtained for the shape of the haloes  must be  taken with caution
although it agrees with previous results. 
Nevertheless, even when the sample can  be limited, results from 
ASOHF  and AHF are completely consistent.

\subsection{Subhaloes}

One of the  main features of the ASOHF finder is  its capacity to deal
with haloes and  subhaloes. In this section we  compare the abundance
and distribution of substructures given by ASOHF and AHF.

For the sake  of comparison, we focus on the  detailed analysis of the
most massive halo in the cosmological simulation previously described.
This halo has a virial  mass of $\sim 8.0 \times10^{14}\, M_\odot$ and
a virial radius  of $\sim 2.4 Mpc$.  To  illustrate the time evolution
of the chosen  halo, we constructed its merger tree  by tracking all its
particles backwards in  time.  In Fig.~\ref{merger_tree} we display the merging
history of the  halo. To facilitate the reading
of this figure, we only show  the mergers among the most massive haloes
that contribute  to build up the  final halo at  $z=0$. Otherwise, the
plot  would  be  saturated by  the  amount  of  mergers due  to  small
structures, which  are not very  relevant from the dynamical  point of
view, though.   The merger tree starts  at $z=0$ and it  plots all the
most massive progenitors of the final halo in previous time-steps over
several output times  of the simulation.  The total  mass of each halo
is represented  by a circle, whose  size is normalized to  the mass of
the  final halo at  $z=0$.  The  meaning of  the different  line types
remains the same as in Sect. 4.3 (Fig. \ref{merger_tree_test3}),
that is the amount of mass  received by any of the progenitors.  This
kind  of plot  not only  shows the  merger history,  but also  the different
interconnections over time.  Although this halo is the most massive in
the simulation, it  is far from being virialized because,  as we can see
in  Fig.~\ref{merger_tree}, it  has suffered  several  major mergers
during its evolution, one  of which happened very recently.  This makes
the process of the substructure analysis more challenging.

\begin{figure*}[!h!t!b!p]
\centering\includegraphics[height=160mm] {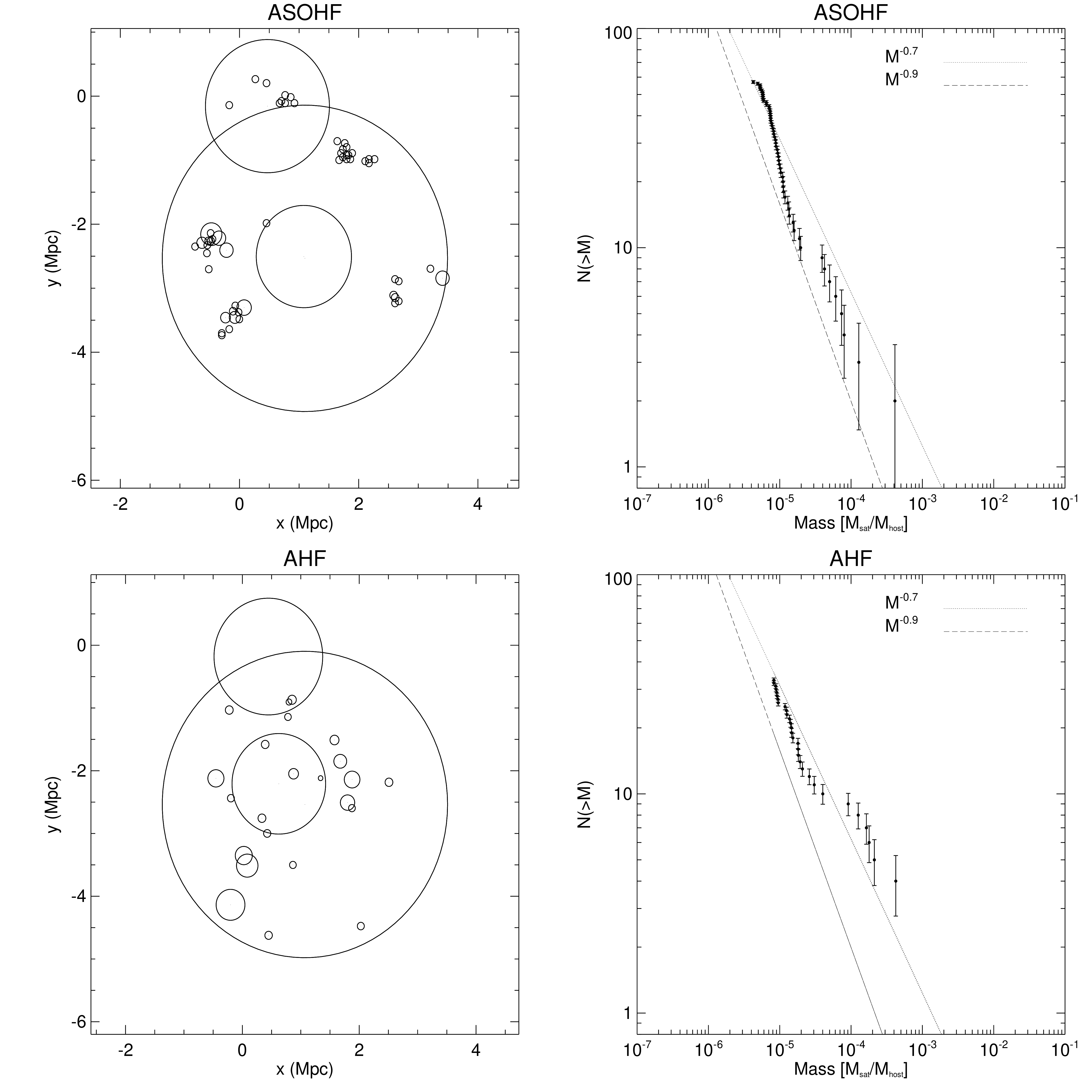}
\caption[Comparison of  substructure.]{Analysis of the subhalo  
population of the most  massive halo  in our  simulation.
Left panels (top and bottom plots):
subhalo  population within the most  massive host halo  in our  simulation
as found by  ASOHF and  AHF. The  size of  the circles
represents the  virial radius of the different  haloes.  Right panels:
cumulative  subhalo mass  function for  the most  massive halo  in our
simulation  as found  by  ASOHF  (upper plot)  and  AHF (lower  plot). 
Two power law fits  with slopes of $-0.7$ and $-0.9$ are
also shown in these panels. Error  bars show $\sqrt N$ uncertainties  
due to the number counts.}
\label{subhaloes}
\end{figure*}

In Fig.~\ref{subhaloes}  we present the analysis  of this particular
halo with  its substructures as found  by ASOHF (upper  plots) and AHF
(lower plots),  respectively.  The left  column of the  panel displays
the 2D  projection of the halo with  its subhaloes.  The x  and y axes
show the  coordinates in  Mpc of the  haloes within  the computational
box. The  comparison of  the haloes identified  by both  codes deserves
some comments.  The  main halo is located at  the same coordinates and
with the  same mass  and size in  both cases.  There is also  a clear
correlation  among the largest  subhaloes in  both subhalo  samples in
sizes and masses. But there seem to be important differences in the
smallest  substructures.   As  we   mentioned  above,  the
explanation of this different performance detecting small structures is
directly  linked with  the  structure  of nested  grids  built by  the
algorithms.

Subhalo  mass functions  have been  widely studied  in  previous works
(e.g.,      \cite{ghigna00};      \cite{delucia04};      \cite{gao04};
\cite{giocoli08};  \cite{Knollmann}), leading  to  the conclusion  that
subhalo mass  functions can  be described with  a power  law, $N_{sub}
(>M) \propto M^{-  \alpha}$, with a logarithmic slope  $\alpha$ in the
range from $0.7$ to $0.9$.  We computed the subhalo mass function
of the  cosmological simulation used  for the comparison of  both halo
finders. The  results are shown in  the panels of the  right column in
Fig.~\ref{subhaloes}.   These  plots   show  the   cumulative  mass
functions of the subhaloes for  the considered main halo as obtained
by  ASOHF  (top)  and  AHF  (bottom).   To  facilitate  the
comparison with previous results, two lines corresponding to the power
laws  with  values  of  $\alpha$   equal  to  $-0.7$  and  $-0.9$  are
plotted. The  masses of  subhaloes are normalized  to the mass  of the
main  halo. The error  bars show  $\sqrt N$  uncertainties due  to the
number counts.

\begin{figure*}
\centering\includegraphics[height=85mm] {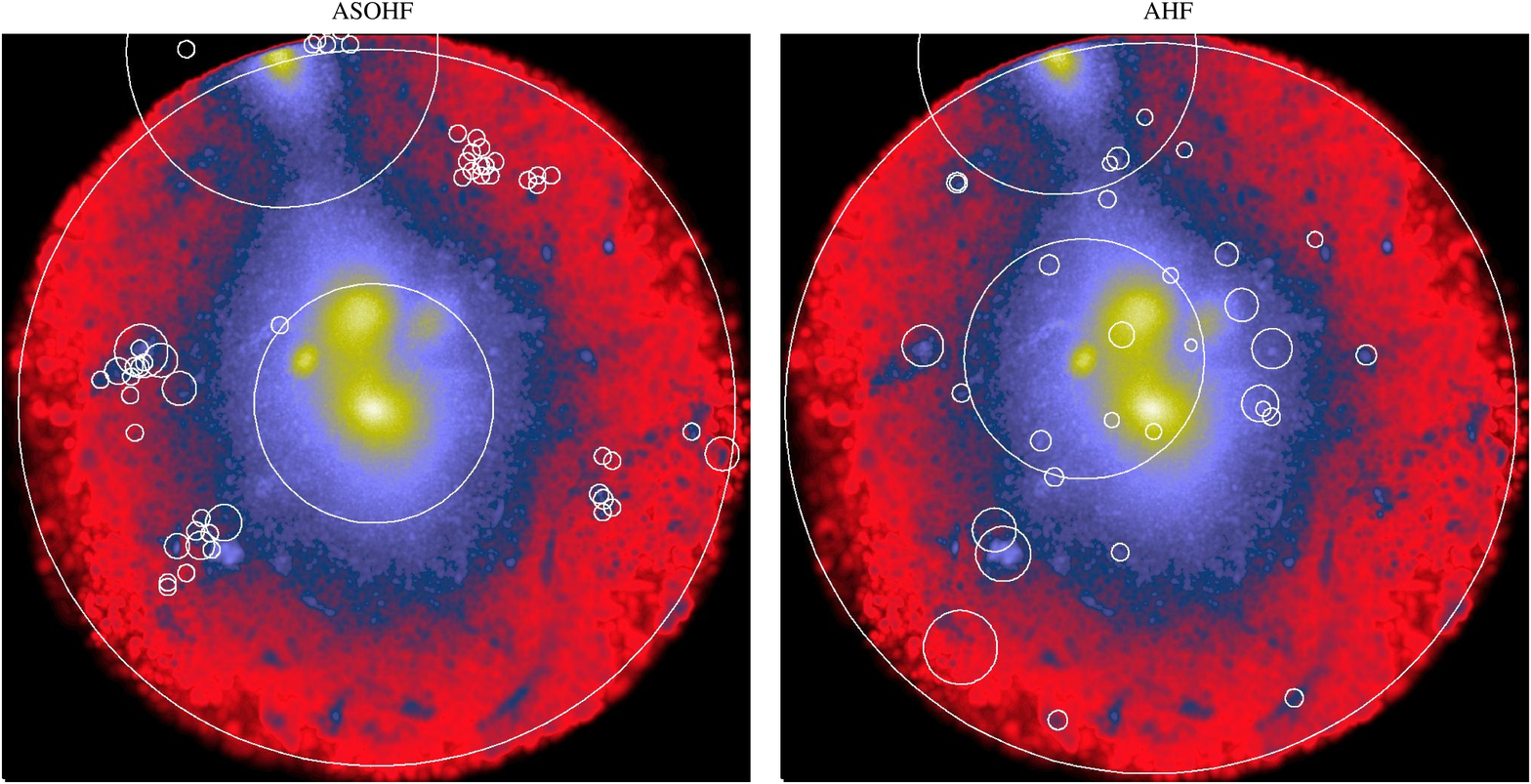}
\caption[Halo and subhaloes]{2D projection  of the dark matter density
field  around  the  biggest  halo  in the  simulation.  The  colour scale
represents the logarithmic of the  density field. Subhaloes  found by
ASOHF (left  panel) and AHF  (right panel) are superimposed  being the
white circles the radii of these haloes.}
\label{halo_color}
\end{figure*}

From  a  statistical  point  of  view, the  comparison  of  both  mass
functions shows that both codes have similar capabilities when dealing
with  finding  substructures.  The  fitting  to  a  power law  of  the
analysed data  gives a  slope of  $0.9$ for ASOHF  and $0.7$  for AHF.  
This would render the ASOHF and AHF mass functions completely
consistent with previous results, because  they are well fit between the two
limiting  power laws  for substructure  in haloes.  Nevertheless, this
conclusion  must be taken  with caution  because, as  we saw, a
direct   comparison  of  the   smallest   substructures   is   not
straightforward.

We  proceed to  compare identified subhaloes  with   the  real  mass  
distribution  in   the  main  halo to  deepen  in  the comparison  of  
the encountered  subhalo
samples and assuming that a direct comparison between both methods is
not  always completely  meaningful, .
Therefore, Fig.~\ref{halo_color}  shows the colour-coded dark matter
column density in  the main halo considered in  this section together
with  the detected subhaloes  overplotted as  circles with  their radii
normalized  to the  main halo  radius. The  left panel  of  the figure
corresponds to the sample obtained  by ASOHF, whereas the right panel
displays  subhaloes  identified by  AHF.   As  we  previously  discussed
when  analysing   Fig.~\ref{subhaloes},  most   of  the
substructures are  unambiguously identified by both codes  and with very
similar features (sizes and  masses).  However, the smallest subhaloes
are not  well identified either with  ASOHF or AHF.  Moreover, it is
striking that  some of these small  substructures do not  match not
only between  both halo  finders, but more  intriguing, with  the real
mass distribution.

\section{Summary and conclusions}

In  the  last years  cosmological  simulations  have experienced  an
astonishing  development,  producing a  huge  amount of  computational
data. Intimately  related to the development of  simulations, all kinds
of analysis tools have arisen  too. One of the most important analysis
tools  have become the  halo finder  algorithms, whose  relevance 
is crucial when comparing simulations with observation.

The  halo-finding  issue  has revealed itself  as  not trivial  at all.   When
cosmological   simulations  have   increased   their  resolution   and
complexity and the  amount of data have grown  exponentially, to find
haloes can  itself be an intensive computational work.  Moreover, the
different  techniques and  implementations  used in  the halo  finders
developed  so far  can  show important  differences, particularly  when
looking at the features of the smallest objects in the simulations.

We developed a new halo-finding code with the main idea of contributing towards 
constraints for a field in which only a  limited number of algorithms are available 
and differences among codes are still relevant.

Our  ASOHF   code  was  especially  designed   to  overcome  some
limitations of the  original SO technique and to  exploit the benefits
of having a  set of nested grids that track  the density distribution in
the analysed volume.  By treating all AMR grids at a certain
level  of refinement  (same numerical  resolution)  independently, the
code is  able to find  haloes at all  levels. 
This procedure can identify  haloes in haloes  in a  natural way 
and therefore describe
the  properties  of the  substructure  in  cosmological objects.   

The
numerical scheme  is also prepared to  compute the merger  tree of the
haloes  in  the  computational  box  as  well  as  some  other  usual
properties of these haloes such as their shapes, and density and velocity
radial profiles.

We set  up several idealised and  controlled tests to check the
capabilities of  the halo  finder.  Although most  of these  tests are
unreal, they allow us total control  of every part of the halo-finding
process. In all tests, the performance of the  finder algorithm has been
correct.

The next step to calibrate and test the ASOHF finder was
to apply  it to the outcome  of a cosmological  simulation and compare
its results with other halo finders widely used like AHF and AFoF.

In a first instance, we compared the sample of haloes encountered
by  the three  codes in  a  given cosmological  simulation.  A  coarse
comparison showed a  good agreement among  them.   The mass
functions obtained  by the three  finders were also  very similar.
We  looked  at the  shape of  the encountered  haloes,  finding a
reasonable concordance  between the results obtained by  ASOHF and AHF
and with previous studies.

More interestingly, we tested the abilities of ASOHF dealing with
substructure  in not idealised  simulations.  In  order to  check this
issue,  we picked up  the most  massive halo  in  the considered
computational box.  This halo was throughly analysed  using both
grid based  on finders, namely,  ASOHF and AHF.  From  the statistical
point  of view,  the  results  are comparable  because  the subhalo  mass
functions are  reasonably consistent. Still,  the comparison object
by object was not as successful because  there are several noticeable
differences concerning  the smaller objects. To clarify this,  
we  compared  the  samples  of  haloes obtained  by  both
algorithms  with the  real mass  distribution. Apparently,  both codes
agree among  themselves and  match the  mass distribution for  the most
relevant features.   However, both algorithms miss  small objects when
they are  compared with the  mass distribution. Surprisingly,  the two
codes do not miss the same small objects.

The explanation for this behaviour for  the ASOHF finder is
related to  how the hierarchy of  nested grids is  created.  We
checked that some of those smaller objects are not always covered by a
high resolution grid.  In that case the halo finder does not identify
the small haloes because it is necessary to have them defined in grids with
enough  numerical resolution.  Although  the detailed  description of
the AHF algorithm is out of  the scope of the present paper, given 
that it  is also a grid  based on halo finder, it  is very likely
that the  differences affecting  the detection of  small substructures
could be caused by the same reasons as in the ASOHF code.

The ASOHF code has been recently used to study galaxy cluster mergers
in a cosmological context  (\cite{planelles09}).  The working version of
the code is  serial and it is written in FORTRAN  95. At present we are
working  on the  parallel  OpenMP version  of  the code,  which will  be
publicly released in due course.

\section*{Acknowledgements}
This work has  been supported by {\it Spanish  Ministerio de Ciencia e
Innovaci\'on}     (MICINN)     (grants    AYA2007-67752-C03-02     and
CONSOLIDER2007-00050)  and  the  {\it Generalitat  Valenciana}  (grant
PROMETEO-2009-103).   SPM thanks  to  the MICINN  for  a FPU  doctoral
fellowship.   The  authors gratefully  acknowledge  S.  Knollmann  and
A. Knebe  for an invaluable help regarding  the use of AHF  as well as
for many  interesting comments and  discussions. The authors also  wish to
thank to  J.M$^{\underline{\mbox{a}}}$.  Ib\'a\~nez and  B. Gibson for
useful discussions, and the anonymous referee for his/her constructive
criticism.   Simulations   were  carried   out  in  the   {\it  Servei
d'Inform\'atica de  la Universitat de Val\`encia} and  the {\it Centre
de Supercomputaci\'o de Catalunya (CESCA)}.



\end{document}